\documentclass[traditabstract]{aa} 
\usepackage{graphicx}
\usepackage{txfonts}
\usepackage{natbib}
\bibpunct{(}{)}{;}{a}{}{,}
\usepackage{float}
\usepackage{calc}
\usepackage{textcomp}
\usepackage{placeins}
\usepackage{color}
\usepackage{rotating} 
\usepackage{lscape}
\usepackage{url}
\usepackage{hyperref}
\usepackage{subfig}
\usepackage[all]{draftcopy}
\usepackage{longtable}
\usepackage{array}
\usepackage{threeparttable}
\usepackage{lscape}
\DeclareTextSymbol{\degre}{OT1}{23}
\newcounter{savedfootnote}

\def \microns{{\,$\mu$m}}
\begin{document}
   \title{Dust Spectral Energy Distributions of Nearby Galaxies: an Insight from the {\it Herschel}\thanks{{\it Herschel} is an ESA space observatory with science instruments provided by European-led Principal Investigator consortia and with important participation from NASA.} Reference Survey. }

\author{L.Ciesla\inst{1,2},
	  M.~Boquien\inst{2,3},
	  A.~Boselli\inst{2},
	  V.~Buat\inst{2},
	  L.~Cortese\inst{4},
	  G.~J.~Bendo\inst{5},
	  S.~Heinis\inst{6},
	  M.~Galametz\inst{7},	  
	  S.~Eales\inst{8},
	  M.~W.~L.~Smith\inst{8},
	  M.~Baes\inst{9},
	  S.~Bianchi\inst{10},
	  I.~de~Looze\inst{9},
	  S.~di~Serego~Alighieri\inst{10},
	  F.~Galliano\inst{11},
	  T.~M.~Hughes\inst{9},
	  S.~C.~Madden\inst{11},
	  D.~Pierini\inst{12},
	  A.~R\'emy-Ruyer\inst{11},
	  L.~Spinoglio\inst{13},
	  M.~Vaccari\inst{14},
	  S.~Viaene\inst{9}
	  C.~Vlahakis\inst{15}.
         }

\institute{	
		University of Crete, Department of Physics, Heraklion 71003, Greece
		\and
		Aix Marseille Universit\'e, CNRS, LAM (Laboratoire d'Astrophysique de Marseille) UMR 7326, 13388, Marseille, France 
		\and
		Institute of Astronomy, University of Cambridge, Madingley Road, Cambridge CB3 0HA, UK
		\and
		Centre for Astrophysics \& Supercomputing, Swinburne University of Technology, Mail H30 - PO Box 218, Hawthorn, VIC 3122, Australia                 	
		\and
		UK ALMA Regional Centre Node, Jodrell Bank Centre for Astrophysics, School of Physics and Astronomy, University of Manchester, Oxford Road, Manchester M13 9PL, United Kingdom
		\and
		Department of Astronomy, University of Maryland, College Park, MD 20742-2421, USA
		\and
		European Southern Observatory, Karl Schwarzschild Str. 2, 85748 Garching bei Muenchen, Germany
		\and
		School of Physics and Astronomy, Cardiff University, Queens Buildings The Parade, Cardiff CF24 3AA, UK
		\and
		Sterrenkundig Observatorium, Universiteit Gent, Krijgslaan 281 S9, B-9000 Gent, Belgium
		\and
		INAF-Osservatorio Astrofisico di Arcetri, Largo Enrico Fermi 5, 50125 Firenze, Italy
		\and
		CEA/DSM/IRFU/Service d'Astrophysique, CEA, Saclay, Orme des Merisiers, Batiment 709, F-91191 Gif-sur-Yvette, France
		\and
		Max-Planck-Institut für extraterrestrische Physik (MPE), Giessenbachstrasse, 85748, Garching, Germany
		\and
		Istituto di Fisica dello Spazio Interplanetario, INAF, Via Fosso del Cavaliere 100, I-00133 Roma, Italy
		\and
		Astrophysics Group, Physics Department, University of the Western Cape, Private Bag X17, 7535, Bellville, Cape Town, South Africa
		\and 
		Joint ALMA Observatory / European Southern Observatory, Alonso de Cordova 3107, Vitacura, Santiago, Chile
}

   \date{Received; accepted}

  \abstract
{
Although accounting only for a small fraction of the baryonic mass, the dust has a profound impact on the physical processes at play in galaxies. 
Thus, to understand the evolution of galaxies, it is essential not only to characterize dust properties per se, but also in relation to global galaxy properties.
   To do so, we derive the dust properties of the galaxies of a volume limited, K-band selected sample, the \textit{Herschel} Reference Survey (HRS).
   
  We gather infrared photometric data from 8\microns\ to 500\microns\ from \textit{Spitzer}, WISE, IRAS and \textit{Herschel} for all of the HRS galaxies.
  Draine \& Li (2007) models are fit to the data from which the stellar contribution has been carefully removed.
  We find that our photometric coverage is sufficient to constrain all of the parameters of Draine \& Li (2007) models and that a strong constraint on the 20-60\microns\ range is mandatory to estimate the relative contribution of the photo-dissociation regions to the infrared spectral energy distribution (SED).
  The SED models tend to systematically under-estimate the observed 500\microns\ flux densities, especially for low mass systems.
  We provide the output parameters for all of the galaxies, i.e. the minimum intensity of the interstellar radiation field, the fraction of polycyclic aromatic hydrocarbon (PAH), the relative contribution of PDR and evolved stellar population to the dust heating, the dust mass and the infrared luminosity.
  For a subsample of gas-rich galaxies, we analyze the relations between these parameters and the main integrated properties of galaxies, such as stellar mass, star formation rate, infrared luminosity, metallicity, H$\alpha$ and H-band surface brightness, and the far-ultraviolet attenuation.
  A good correlation between the fraction of PAH and the metallicity is found implying a weakening of the PAH emission in galaxies with low metallicities and, thus, low stellar masses.
  The intensity of the diffuse interstellar radiation field and the H-band and H$\alpha$ surface brightnesses are correlated, suggesting that the diffuse dust component is heated by both the young stars in star forming regions and the diffuse evolved population.
  
  We use these results to provide a new set of infrared templates calibrated with \textit{Herschel} observations on nearby galaxies and a mean SED template to provide the $z$=0 reference for cosmological studies.
  For the same purpose, we put our sample on the $SFR$-$M_*$ diagram.
  The templates are compared to the most popular infrared SED libraries, enlightening a large discrepancy between all of them in the 20-100\microns\ range.
}

  
   \authorrunning{Ciesla et al.}
   \titlerunning{SED fitting of the \textit{Herschel} Reference Survey}

   \maketitle

\section{\label{intro}Introduction}

In the interstellar medium (ISM), dust holds a major role: 
a) it acts as a catalyst in the transformation of the atomic hydrogen into molecular hydrogen from which stars form \citep{Wolfire95},
b) it allows the gas to cool and condense to form new stars by absorbing the ultraviolet (UV) emission of surrounding young stars \citep{Draine78,Dwek86,Hollenbach&Tielens97},
c) it re-emits the absorbed energy in the infrared domain (IR), where the thermal emission by dust grains dominates the spectral energy distribution (SED) of galaxies between $\approx$10 and 1000\microns.
Dust is thus an important tracer of the star formation activity.
Formed by the aggregation of metals injected into the ISM by stars through stellar winds and supernovae explosions, its composition still remains uncertain.
The most popular models assume that dust consists of a mixture of silicate and graphite grains \citep{Mathis77,DraineLee84,Kim94}, and are extended by adding the contribution of the polycyclic aromatic hydrocarbons (PAH), producing the broad spectral features seen in the mid-IR (MIR), such as in \cite{SiebenmorgenKrugel92}, \cite{LiDraine01}, \cite{WeingartnerDraine01} and \cite{DraineLi07} models.

Because of the important role of dust in the ISM and its tight link with the other components of galaxies, the study of dust emission is mandatory to have a better understanding of all of the processes at play.
IRAS (Neugebauer et al. 1984), COBE (1989), ISO (Kessler et al. 1996), \textit{Spitzer} (Werner et al. 2004), and AKARI \citep{Murakami07} allowed us to study dust emission up to 240~$\mu$m. 
While these telescopes sampled well the dust properties from the MIR to around the peak of the dust SED, going to longer submillimeter (submm) wavelengths, well beyond the peak, is crucial to model the distribution of the dust size, temperature and composition. 
In this way  the inventory of the bulk of the dust mass in galaxies, manifested in the submm, is also not missed \citep{DevereuxYoung90,Gordon10}.
Ground-based facilities, such as SCUBA on JCMT \citep{Holland99}, provide us with submm data but observations of large samples of normal galaxies are still prohibitive due to the long integration times needed for these instruments and suffer from a limited sensitivity.
Furthermore, a large part of the far-IR (FIR) and submm domains are not accessible from the ground.
The \textit{Herschel} Space Observatory (Pilbratt et al. 2010) opened a new window on the far-IR/submm spectral domain (55 to 672\microns) allowing us to probe the cold dust component in a large number of nearby objects.

We now have a global view of the emission from all of thermal dust components, covering a broad range of temperatures, and the different heating mechanisms for these components.  
Before \textit{Herschel}, it was already evident that a significant part of the IR spectral energy distribution (SEDs) of galaxies potentially includes cold (17 to 20\,K) diffuse cirrus components heated by the diffuse interstellar radiation field from the total stellar population, and not just the light from star forming regions \citep[e.g.][]{Helou86,XuHelou96,LiDraine02,Boselli04,Komugi11}.  
However, the emission for these colder components often appears intermixed with emission from warmer dust heated by star forming regions at $\lambda$ lower than $\approx$240\microns, and in many cases, the emission is poorly constrained.  
Multiple authors using \textit{Herschel}, including \cite{Bendo10,Bendo12a}, \cite{Boquien11}, \cite{Groves12}, and \cite{Smith12}, have used comparisons of FIR colors to NIR emission from the evolved stellar population to demonstrate that the cirrus component is often the primary source of $>$250\microns\ emission and is contributing also at shorter wavelengths. 

Understanding the heating processes of the dust is of paramount importance in order to provide physical models than can reproduce the FIR/submm SED of galaxies.
Multiwavelength radiative transfer modeling of galaxies is a powerful tool to analyze the properties of dust in galaxies in a self-consistent way \citep[e.g.][]{Xilouris99,Popescu00,Alton04,Bianchi08,Baes10,MacLachlan11,Popescu11,DeLooze12a,DeLooze12b}.
However, assumptions on the geometry of the dust need to be made, and these codes require large computational resources, especially for large sample of galaxies.
Therefore, models, as well as empirical templates, are widely used to extract information about galaxies from their IR SED \citep[][hereafter CE01, DH02, B03, and DL07]{CharyElbaz01,DaleHelou02,Boselli03a,DraineLi07}.
The pre-\textit{Herschel }empirical libraries of templates (CE01, DH02) were calibrated on FIR observations and detections of galaxies up to $\approx$200\microns, and were constructed from local normal and IR luminous star forming galaxies.
 Recent studies, making use of the new \textit{Herschel} data, provided new IR templates for low and high redshift objects \citep{Elbaz11,SmithDunne12,Magdis12}.
However, there is a lack of templates representative of the broad variety of nearby normal galaxies, reference of the $z$=0 Universe.

The \textit{Herschel} Reference Survey \citep[HRS,][]{Boselli10a} is composed of 322 nearby galaxies spanning the entire range of morphological types and environment.
The aim of this volume-limited, K-band selected, sample is to investigate the dust properties of galaxies for which a wealth of ancillary data are available, both photometric \citep[from UV to radio,][Cortese et al., submitted]{Bendo12b,Boselli10a,Boselli11,Ciesla12,Cortese12a} as well as spectroscopic \citep[][Boselli et al., in press]{Boselli13,Hughes13}.
With this set of data, the HRS is one of the best samples to study the dust properties of nearby galaxies versus parameters such as the stellar mass, the metallicity, the star formation rate, etc, giving us a better understanding of its role in the ISM and the interplay with the stellar radiation field.

In this work, we derive the physical properties of the dust by fitting the integrated IR SED of the HRS galaxies with the models of DL07.
Although these models are physical, they have been mainly tested on FIR data \citep{Draine07}, and recently, using \textit{Herschel} data \citep{Dale12,Aniano12,Draine14}.
We thus discuss the ability of those models to reproduce the submm observations of our galaxies. 
For a subsample of gas-rich galaxies, we then investigate the relations between the derived output parameters and the integrated galaxy properties (stellar mass, star formation rate, birthrate parameter, metallicity, FUV attenuation, H$\alpha$ surface brightness and H-band effective surface brightness).
Based on this analysis, we derive a set of SED templates, binned according to those parameters that better characterize the shapes of the SED, and compare them with libraries available in the literature.

This paper is organized as follows: in Section~\ref{sample}, we describe the HRS and define a gas-rich subsample, on which we will focus our analysis.
The ancillary data associated to this subsample are presented in Section~\ref{data}.
We present the procedure used to perform the fitting, analyze our ability to constrain the models with the data and present the results in Section~\ref{sedfitting}.
For the gas-rich galaxy subsample, we outline the most interesting relations between the output from the fitting and galaxy parameters in Section~\ref{comp}.
The derivation and discussion of the new IR library is presented in Section~\ref{temp}.
In Appendix~\ref{photom}, we present the MIR photometry of all of the HRS galaxies at 8, 12 and 22\microns, which are important constraints for the SED modeling.
In Appendix~\ref{stell}, we explain the method derived to remove the stellar contribution from our data and provide new coefficients useful for future works.
For completeness, in Appendix~\ref{defparam}, we give the output parameters from the SED fitting for the galaxies which are not discussed in the main part of this work (early-type galaxies and H{\sc i}-deficient galaxies).
Finally, all of the relations between the output of DL07 models and the properties of the gas-rich sample are presented and discussed in Appendix~\ref{apcompparam}.
\section{\label{sample}The Sample} 

The HRS galaxies are selected according to three criteria fully described in \cite{Boselli10a}.
The HRS is a volume limited sample composed of galaxies lying at a distance between 15 and 25\,Mpc.
The galaxies are then selected according to their K-band magnitude, the luminosity of which is a proxy for the total stellar mass \citep{Gavazzi96}. 
Based on optical extinction studies and FIR observations at wavelengths shorter than 200~$\mu$m, we expect late-type galaxies to have a larger content of dust than early-types \citep{Sauvage94}.
Thus, two different $K_{mag}$ limits have been adopted: $K_{mag}\leq$12 for late-types and $K_{mag}\leq$8.7 for early-types (in Vega magnitudes).
Finally, to limit the contamination from Galactic cirrus, galaxies are selected at high Galactic latitude ($b > + 55\deg$) and in low Galactic extinction regions \citep[$A_{B} < 0.2$,][]{Schlegel98}.
The final sample contains 322\footnote{With respect to the original sample given in \cite{Boselli10a}, the galaxy HRS\,228 is removed from the complete sample because its updated redshift on NED indicates it as a background object.}  galaxies, among which 62 early-types and 260 late-types.
The HRS covers all morphological types and contains field galaxies as well as objects in high density regions such as the Virgo cluster.
We use the morphological classification presented in \cite{Cortese12b}.

Even if we present the NIR photometry and perform the SED fitting of all of the HRS galaxies, the paper focusses on the study of a subsample of 146 gas-rich galaxies.
From the whole HRS sample, we remove early-type galaxies (E-S0-S0/Sa) from the analysis as the dust properties and the dust heating sources of elliptical and lenticular galaxies are different from late-type galaxies \cite[e.g.][]{Boselli10b,Smith12}.
Indeed, the relative contribution of X-ray heating, stochastic heating, heating from fast electrons in the hot gas, and the size-distribution of dust grains in these environments with low-density ISM might differ from that of late-types and thus need further investigations \citep{Wolfire95}.
Furthermore, only 32\% of the elliptical galaxies and 60\% of the lenticulars are detected at 250\microns\ \citep{Ciesla12}, yielding to an incomplete photometric coverage of the IR-submm domain.
Finally, some of the physical properties used in this work (birthrate parameter, H$\alpha$ surface brightness and metallicity) are not available for all of the early-type galaxies.
Furthermore, a number of late-type galaxies of the HRS lie in the very dense environment of the Virgo cluster.
These galaxies have their gas content stripped by the environment \citep{BoselliGavazzi06}.
\textit{Herschel} observations have recently shown that the dust component of H{\sc i}-deficient\footnote{The H{\sc i}-deficiency,  $H{\sc i}-def$, is defined as the difference, in logarithmic scale, between the HI mass expected from an isolated galaxy with the same morphological type and optical diameter and the observed HI mass \citep{Haynes84}.} galaxies is also affected by the cluster environment \citep{Cortese10b,Cortese12a}.
In the following, we define as ``gas-rich'' galaxies those with H{\sc i}$-def \leq 0.4$, and ``H{\sc i}-deficient'' those with H{\sc i}$-def > 0.4$, to be consistent with \cite{Boselli12}.
We decide to not consider H{\sc i}-deficient galaxies in our analysis to remove the effects of the environment as a free parameter that could bias the interpretation of the results.
The study of the effect of the environment on the dust properties of galaxies will be presented in a future work.

The main part of this paper is focused on the data, the SED fitting, and the analysis of the gas-rich late-type sample. The NIR photometry of all of the HRS galaxies, and the output parameters from the SED fitting obtained for the early-type galaxies and the H{\sc i}-deficient galaxies are presented in Appendix~\ref{photom} and Appendix~\ref{defparam}, respectively.

\section{\label{data}Data} 
	
	In order to compute the IR SEDs of the HRS galaxies, we use data from 8 to 500\microns\ performing the photometry on \textit{Spitzer}/IRAC and WISE images and using measurements available in the literature from \textit{Spitzer}/MIPS, \textit{Herschel}/PACS, \textit{Herschel}/SPIRE, and IRAS.
	These data are publicly available through the Hedam database \footnote{\url{http://hedam.lam.fr/HRS/}}.

	\subsection{Mid-infrared: \textit{Spitzer}/IRAC and \textit{WISE}}

	For the purpose of this work, we perform the photometry of 56 out of the 146 galaxies of our subsample for which observations at 8\microns\ from \textit{Spitzer}/IRAC were available in the \textit{Spitzer} archive.
	The procedure is presented in Appendix~\ref{photom} and the flux densities and associated errors are given in Table~\ref{MIRflux}.
	Our procedure to estimate the errors associated to the 8\microns\ data, based on \cite{Boselli03a}, results in a mean error of 15\% (see Table~\ref{phot_compl}).
	We also perform the WISE photometry of the gas-rich galaxies at 12 and 22\microns.
	As for the IRAC 8\microns, the procedure used, the flux densities and errors associated are fully described in Appendix~\ref{photom}.
	The mean errors associated to the 12\microns\ and the 22\microns\ flux densities are 6\% and 13\%, respectively. 
	Comparisons between our measurements and results from the literature, as presented in Appendix~\ref{photom}, are in good agreement.
	IRAS 12\microns\ flux densities are available only for 15\% of our galaxies, whereas all of our subsample galaxies have a 12\,$\mu$m flux density from WISE.
	In order to have an homogeneous set of data, we use the WISE 12\microns\ flux densities.

	\subsection{\label{fir}Far-infrared: \textit{Spitzer}/MIPS, IRAS, and \textit{Herschel}/PACS}
	
	The reduction and photometry of MIPS data are fully described in \cite{Bendo12b}.
	Flux densities of 68 gas-rich galaxies are available at 24\microns, and 47 at 70\microns.
	For most of the galaxies, aperture photometry was performed using an elliptical region with major and minor axes of 1.5 times the axis sizes of the $D_{25}$ isophotes given by \cite{DeVaucouleurs91}.
	The same aperture has been used in the two bands.
	In the case of HRS~20-NGC~3395 and its companion NGC~3396, the flux densities provided correspond to one of the pair as it is hard to disentangle the emission from the two galaxies within a pair.
	Thus, we do not use these measurements.
	The error calculation takes into account the calibration error, 4 and 10\% at 24 and 70\microns\ \citep{Engelbracht07,Gordon07}, respectively, the uncertainty based on the error map, and the background noise.
	The three added in quadrature.
	We should note that a transcription error was made in \cite{Bendo12a} for the flux density of HRS\,142.
	Its flux density at 70\microns\ is 8200\,mJy instead of 1237\,mJy.
	Despite the incompleteness of the MIPS 24\microns\ data, we choose to use them, when available, instead of the WISE 22\microns.
	This choice is due to the poor quality of some WISE 22\microns\ images resulting in a mean error of 13\% in this band when the mean error of MIPS 24\microns\ is 4\% only.
	IRAS 25\microns\ data are also available but for a small part of our subsample (15\%), we thus decided to not use these measurements to keep an homogeneous set of data from galaxy to another.

	Because of the incompleteness at 70\microns, we also use, when available, the 60\microns\ measurements from IRAS presented in \cite{Boselli10a}.
	IRAS 60\microns\ flux densities of 128 galaxies of the subsample, with a typical uncertainty of 15\%, are provided by multiple references in the literature and collected on NED: \cite{Sanders03,Moshir90,ThuanSauvage92,Soifer89,Young96}.

	All of the gas-rich galaxies are detected by PACS at 100 and 160\microns, respectively (Cortese et al., in press).
	They performed aperture photometry following the method used by \cite{Ciesla12} to build the HRS SPIRE catalog (see Section~\ref{spirephot}).
	The apertures used to extract the fluxes are identical to those used for the SPIRE photometry.
	However, a refinement of the photometric aperture has been applied to some of the objects for various reasons (some galaxies were unresolved in SPIRE bands and resolved in PACS bands, some PACS maps were too small to encompass the SPIRE aperture, or the FIR emission of the galaxy was much less extended than the SPIRE aperture).
	Errors were estimated following the method described in \cite{Roussel13}.
	The mean errors are 16\% and 12\% at 100 and 160\microns, respectively.
	The data reduction and the integrated photometry are described in Cortese et al., submitted.

	\subsection{\label{spirephot}Submillimetre: \textit{Herschel}/SPIRE}		
	
	In the SPIRE bands, all of the galaxies of the subsample are detected at 250, 350, and 500\microns, respectively.
	The SPIRE photometry is fully described in \cite{Ciesla12}.
	In summary, aperture photometry was performed in elliptical regions for extended galaxies.
	All apertures have been chosen to encompass the emission and to minimize the contamination of background sources.
	The stochastic error takes into account the instrumental uncertainty, the confusion uncertainty (due to the presence of faint background sources), and the background uncertainty (due to large scale structure such as cirrus), all three added in quadrature.
	Flux densities of point-like sources have been measured using PSF fitting on timeline data \citep{Bendo13}.
	We take into account the last updates by applying the calibration corrections (1.0253, 1.0250 and 1.0125 at 250, 350, and 500\microns, respectively) and new beam areas\footnote{See \url{http://herschel.esac.esa.int/twiki/bin/view/Public/SpirePhotometerBeamProfileAnalysis}} of 450, 795 and 1665\,arcsec$^2$ at 250, 350, and 500\microns, respectively.
	We do not take into account the variations of the beam sizes depending on the shape of the SED as they are generally within the SPIRE errors.
	The result of these updates lowers the flux densities presented in \cite{Ciesla12} by $\approx$5\%. 
	The mean errors are 6\%, 8\% and 11\% at 250, 350, and 500\microns, respectively.
	
	The photometric completeness of the gas-rich sample of galaxies is presented in Table~\ref{phot_compl}.	
	Upper limits are not taken into account in our fitting procedure and will thus correspond to an absence of data in Table~\ref{phot_compl}.

	\begin{table}
		\centering
		\caption{Completeness of the photometric coverage of the 146 gas-rich galaxies. Only detections are considered. }
		\begin{tabular}{l c c c}
	  	\hline\hline
		Band & $\lambda$ (\microns) & Mean error (\%) & Number of galaxies\\ 
		\hline		
		\textit{Spitzer}/IRAC			&8		&	15  & 56 \\			
		WISE						&12		&	6	& 146\\
		WISE						&22		&	13	& 141\\
	  	\textit{Spitzer}/MIPS			&24		&	 4	& 68\\
		IRAS						&60		&	 15	& 128\\
		\textit{Spitzer}/MIPS			&70		&	 10	& 47\\
		\textit{Herschel}/PACS		&100	&	 16	& 146\\
		\textit{Herschel}/PACS		&160	&	 12	& 146 \\
		\textit{Herschel}/SPIRE		&250	&	 6	& 146\\
		\textit{Herschel}/SPIRE		&350	&	 8	& 146\\
		\textit{Herschel}/SPIRE		&500	&	 11	& 146\\
		\hline
		\label{phot_compl}
		\end{tabular}
	\end{table}
	\section{\label{sedfitting}SED fitting with the \cite{DraineLi07} models}

		\subsection{\cite{DraineLi07} models}
		
		DL07 modeled the dust with a mixture of astronomical amorphous silicate and carbonaceous grains with the size distribution observed in the Milky Way \citep{WeingartnerDraine01}.
		Models of the Large and Small Magellanic Cloud grain size distribution are also available.		
		
		The bulk of the dust present in the diffuse ISM is heated by a large number of stars responsible of the diffuse radiation.
		However, another part of the dust is located in regions close to very luminous O and B stars, in photo-dissociation regions (PDR).
		In PDRs, the light coming from the young stars is heating the dust and is much more intense than the emission coming from the old stars responsible of the diffuse radiation.
		In DL07 models, the relative dust mass fraction heated by each source, the diffuse component and the PDRs, is given by the $\gamma$ parameter.
		Thus, the fraction $(1-\gamma)$ of the total dust mass is heated by $U=U_{min}$, with U the intensity of the interstellar radiation field (ISRF) and $U_{min}$ the intensity of the diffuse ISRF, both normalized to the intensity of the Milky Way ISRF.
		Recently, \cite{Aniano12} linked the temperature of the cold dust component to the $U_{min}$ parameter by approximating the DL07 SED with a blackbody multiplied by a power-law opacity and obtained $T_d\approx$20$U_{min}^{0.15}$K.
		The fraction $\gamma$ of the total dust mass is exposed to a range of stellar intensities following a power law distribution from $U_{min}$ to $U_{max}$ with $dM/dU \propto U^{-2}$.
		From these 3 parameters, we can compute $<U>$, the mean intensity of the ISRF, from equation 17 of \cite{Draine07}.
		A last parameter aims at characterizing the emission due to the PAH. 		
		Their abundance is quantified with the parameter $q_{PAH}$ which corresponds to the fraction of the total grain mass contributed by PAH containing less than 10$^3$\,C atoms.
		Each model depends on the set of parameters \{$dust composition$, $q_{PAH}$, $\gamma$, $U_{min}$, and $U_{max}$\}.
		Finally, the dust mass is also a free parameter determined from the normalization of the model to the observations.
		Following the recommendations of \cite{Draine07}, we use only the Milky Way dust type and fix $U_{max}=10^6$.
		Thus the free parameters of the fit are $q_{PAH}$, $\gamma$, and $U_{min}$.
		The normalization of the model to the data provides us with the dust mass $M_{dust}$, and the integration of the model between 8 and 1000\microns\ gives us with the infrared luminosity $L_{IR}$.

		\subsection{\label{fitproc}Fitting procedure}
		
		As DL07 model the emission of the dust from 1\microns\ to 1\,cm, we need to remove the stellar emission that contributes to the MIR data in order to have only the emission from the dust.
		In Appendix~\ref{stell}, we describe the method we use to remove the stellar contribution and provide morphological type dependent coefficients, normalized to several NIR bands, determined using the CIGALE\footnote{\url{http://cigale.lam.fr/}} code \citep[Code Investigating GALaxy Emission,][]{Noll09}.
		The code computes modeled galaxy SEDs by using stellar population models from \cite{Maraston05} which are convolved with a given star formation history (SFH).
		We use an exponentially decreasing SFH as we present in Appendix~\ref{stell}. 
		The coefficients ($S_{\nu,stellar}/S_{\nu}$) applied to the MIR photometry of the sample are 0.093, 0.075, 0.016 and 0.017 at 8, 12, 22 and 24\microns, respectively (see Appendix~\ref{stell}) .		

		The DL07 models are integrated into the filters of the corresponding photometric bands, and these modeled flux densities are compared to the observations.
		The ranges allowed for each parameter are presented in Table~\ref{rangeparam}.
		For galaxies that do not have any MIPS 24\microns\ observations, the fit is performed using the WISE 22\microns\ flux densities as explained in Section~\ref{fir}.
		
		\begin{table}
			\centering
			\caption{Ranges of parameter used in our fitting procedure. }
			\begin{tabular}{l c}
	  		\hline\hline
			Parameter & Values\\ 
			\hline		
			$U_{min}$ 			&	from 0.10 to 25		 \\
			$U_{max}$			&	$10^6$	\\
			$q_{PAH}$ ($\%$)	& 	from 0.47 to 4.58		\\
			$\gamma$ ($\%$)	&  100 logarithmically-spaced values from 0.1\% to 100\% \\
			\hline
			$L_{IR}$			& Integration of the model between 8 and 1000\microns	\\
			$M_{dust}$		& Normalization of the model to the data	\\
			\hline
			\label{rangeparam}
			\end{tabular}
		\end{table}
		
		We use a  $\chi^2$ minimization method to fit our data with the models of DL07.	
		For each galaxy, we compute the $\chi^2$ corresponding to every model using the equation:
		\begin{equation}
		\chi^2 (a_1,...,a_i,...,a_N)= \sum_{i=1}^{M}\left[\frac{y_i - \alpha y(x_i,a_1,...,a_i,...,a_N)}{\sigma_i}\right]^2,
		\end{equation}
		and the reduced $\chi^2$ as:
		\begin{equation}
		\chi^2_{red} = \frac{\chi^2}{M-N}
		\end{equation}
		\noindent
		where $y$ is the model, $a_i$ are the parameters values of this model, $x_i$ correspond to the observations, and $\sigma_i$ the errors attributed to these observations.
		$N$ is the number of parameters and $M$ the number of observed data.
		We obtain the normalization factor through the following equation:

		\begin{equation}
		\alpha = \frac{\sum_{i=1}^{N}y_i \times y(x_i,a_1,...,a_i,...,a_M)/\sigma_i^2}{\sum_{i=1}^{N}y(x_i,a_1,...,a_i,...,a_M)^2/\sigma_i^2}.
		\end{equation}

		As the reduced $\chi^2$ ($\chi^2_{red}$) is calculated for each value of a parameter, we can build the probability distribution function (PDF) of this parameter.
		For each value of a discrete parameter, we select the corresponding minimum $\chi^2_{red}$.
		Thus we have the distribution of the minimum $\chi^2_{red}$ associated to the set of values of the parameter.
		From this distribution, we obtain the estimated value of the parameter, as the mean value of this distribution, and the error associated as its standard deviation, proceeding as described in \cite{Noll09}.
		For parameters with a large range of values, we compute bins and take the minimum $\chi^2_{red}$ corresponding to each bin.
		Then we build the distribution.
		In the following discussions, we will refer to ``best'' parameter as the parameter obtained directly from the model providing the minimum $\chi^2_{red}$, and to ``estimated'' parameter as the parameter derived from the distribution of $\chi^2_{red}$.
		
		This SED fitting procedure is applied to all of the galaxies of the HRS.
		However, from now on, this work only focus on the gas-rich galaxy subsample and the results for the other galaxies (early-type, and late-type deficient galaxies) are presented in Appendix~\ref{defparam}.

		\subsection{\label{mocks} Mock catalogs}		

		The $\chi^2$ fitting described in the previous section provides us with a best fit model for each galaxy.
		However, we need to know if the output parameters obtained from these models are reliable, i.e. if the data we have allow us to constrain these parameters.
		To do so, we create mock catalogs following the procedure outlined in \cite{Giovannoli11}.
		We first run our $\chi^2$ procedure on our sample in order to obtain the best-fitting model for each galaxy and the corresponding parameters.
		The resulting best SEDs are integrated in the filters adding an error randomly distributed according to a Gaussian curve.
		The $\sigma$ of the Gaussian curve is chosen to be the median value of the error for each band.
		We now have a mock catalog from which we know the exact parameters associated with.
		As a final step, we run our fitting procedure on this mock catalog and compare the input ``best'' values of the parameters to the output ``estimated'' values.
		This test allows us to evaluate the ability to constrain a parameter with the photometric coverage available for our galaxies.

		 \begin{figure}
			\includegraphics[width=9cm]{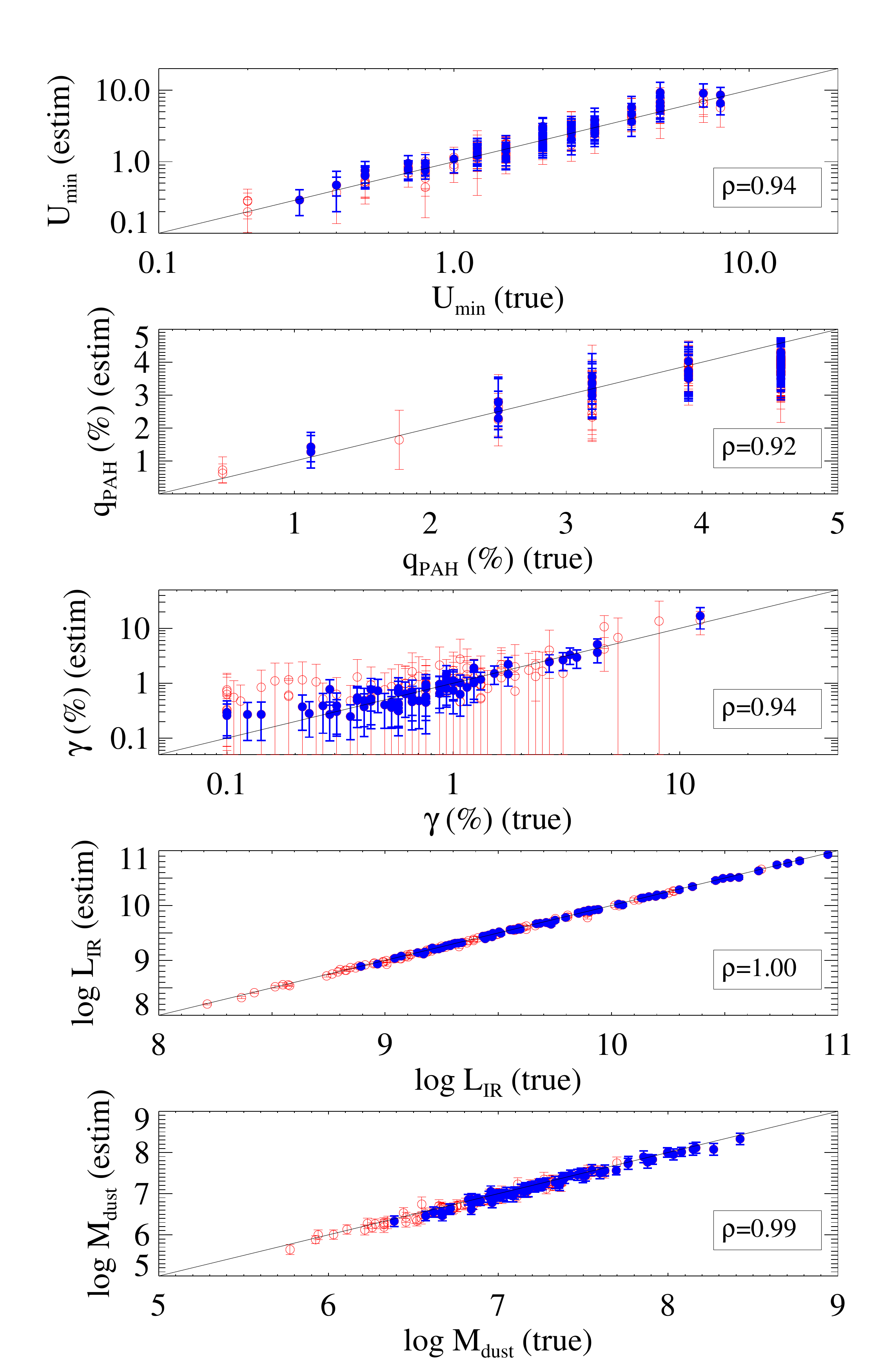}
  			\caption{ \label{mock1}Comparison between the true values of the parameters, from top to bottom: $U_{min}$, $q_{PAH}$, $\gamma$, $\log L_{IR}$, $\log M_{dust}$ and the mock catalogs parameters values estimated from the PDF. Blue filled dots correspond to galaxies having a 24\microns\ measurement and red empty dots are galaxies with a 22\microns\ measurement. The solid black line is the 1:1 relationship. DL07 parameters are constrained, but $\gamma$ needs the 24\microns\ from MIPS to be properly estimated. }
		\end{figure}
				
		 \begin{figure*}
			\includegraphics[width=\textwidth]{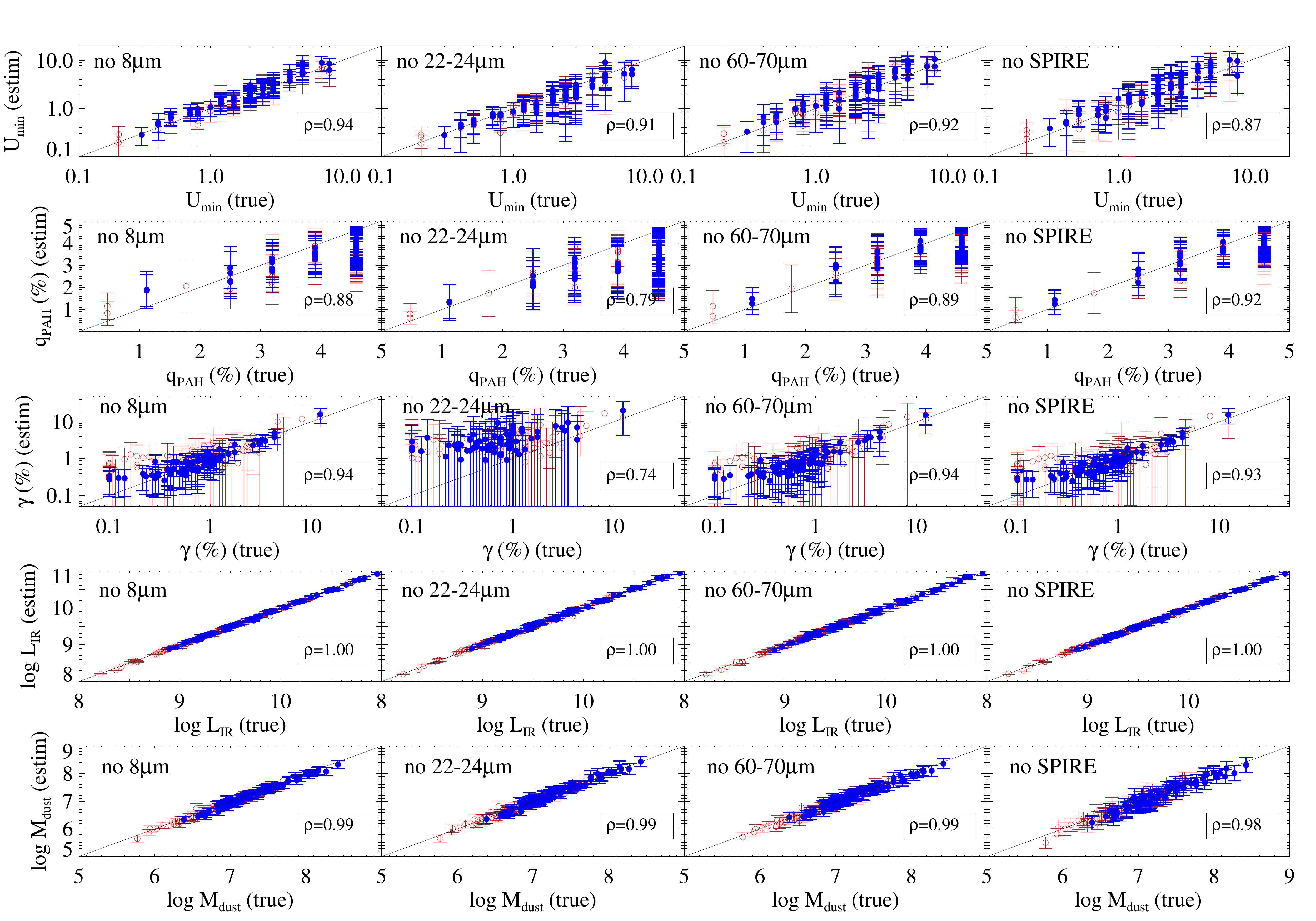}\hfill %
  			\caption{ \label{mock2}Comparison between the true values of the parameters (from top to bottom) $U_{min}$, $q_{PAH}$, $\gamma$, $\log L_{IR}$, $\log M_{dust}$ and the mock catalogs parameters values estimated from the PDF. The different panels correspond to different observed band combinations. For each parameter and from the left to the right: all bands but the 8\microns, all bands but the 22 and 24\microns, all bands but the 60 and 70\microns, and all bands but SPIRE. Blue filled dots correspond to galaxies having a 24\microns\ measurement and red empty dots are galaxies with a 22\microns\ measurement. The solid black line is the 1:1 relationship. DL07 parameters are constrained, but $\gamma$ needs the 24\microns\ from MIPS to be properly estimated. }
		\end{figure*}

		\begin{table*}
			\centering
			\caption{Statistics from the relations of the mock catalogs (Figure~\ref{mock1}).}
			\begin{tabular}{l l c c c c c }
			\hline\hline
	  		\multicolumn{1}{c}{Parameter} 	& \multicolumn{1}{c}{Bands} & \multicolumn{2}{c}{Linear fit: A + B$x$}	& \multicolumn{1}{c}{Spearman coef.}& \multicolumn{2}{c}{Stat of $X_{est}/X_{true}$} \\  
			\hline
			\multicolumn{1}{c}{} 	& \multicolumn{1}{c}{} & \multicolumn{1}{c}{A} & \multicolumn{1}{c}{B}	& \multicolumn{1}{c}{$\rho$}& \multicolumn{1}{c}{Median} & \multicolumn{1}{c}{$\sigma$} \\			\hline		
			$U_{min}$	&	all					& 0.03 & 1.07 & 0.94 & 1.07 & 0.20 \\			
						&	no 8\microns\		& 0.03 & 1.07 & 0.94 & 1.06 & 0.20 \\
						&	no 22-24\microns\	& 0.08 & 0.81 & 0.91 & 0.86 & 0.19 \\
	  					&	no 60-70\microns\ 	& 0.07 & 1.18 & 0.92 & 1.18 & 0.30 \\
						&	no SPIRE			& 0.21 & 1.15 & 0.87 & 1.24 & 0.34 \\
			\hline
			$\gamma$	&	all					& 0.00 & 1.25 & 0.94 & 1.18 & 1.36 \\ 
						&	no 8\microns\		& 0.00 & 1.17 & 0.94 & 1.19 & 1.43 \\
						&	no 22-24\microns\	& 0.02 & 1.25 & 0.74 & 3.77 & 6.43 \\
						&	no 60-70\microns\ 	& 0.00 & 1.14 & 0.94 & 1.23 & 1.36 \\
						&	no SPIRE			& 0.00 & 1.10 & 0.93 & 1.17 & 1.33 \\
			\hline																
			$q_{PAH}$	&	all					& 0.74 & 0.71 & 0.92 & 0.89 & 0.11 \\
						&	no 8\microns\		& 1.00 & 0.61 & 0.88 & 0.85 & 0.20 \\
						&	no 22-24\microns\	& 0.88 & 0.55 & 0.79 & 0.77 & 0.15 \\
						&	no 60-70\microns\ 	& 1.01 & 0.63 & 0.89 & 0.89 & 0.17 \\
						&	no SPIRE			& 0.84 & 0.69 & 0.92 & 0.89 & 0.14 \\
			\hline	
			$\log L_{IR}$	&	all					& 0.00 & 1.00 & 1.00 & 1.00 & 0.00 \\
						&	no 8\microns\		& 0.01 & 1.00 & 1.00 & 1.00 & 0.00 \\
						&	no 22-24\microns\	& -0.03 & 1.00 & 1.00 & 1.00 & 0.00 \\
						&	no 60-70\microns\ 	& -0.09 & 1.01 & 1.00 & 1.00 & 0.00 \\
						&	no SPIRE			& -0.02 & 1.00 & 1.00 & 1.00 & 0.00 \\
			\hline	
			$\log M_{dust}$	&	all					& -0.03 & 1.00 & 0.99 & 0.99 & 0.01 \\
							&	no 8\microns\		& -0.04 & 1.00 & 0.99 & 0.99 & 0.01 \\
							&	no 22-24\microns\	& -0.05 & 1.01 & 1.01 & 1.00 & 0.01 \\
							&	no 60-70\microns\ 	& 0.06 & 0.98 & 0.98 & 0.99 & 0.01 \\
							&	no SPIRE			& -0.14 & 1.01 & 1.01 & 0.99 & 0.02 \\				
			\hline
			\label{statmock}
			\end{tabular}
		\end{table*}

		We note that some galaxies of our sample are lacking 8\microns, 22-24 or 60-70\microns\ measurements.
		In order to estimate the impact of the lack of data on the estimation of the parameters, we build several mock catalogs using different combinations of bands to understand the importance of every photometric band.
		Indeed, all of the galaxies of our sample were not observed in every band.
		Besides understanding the effect of the inhomogeneity of the photometric coverage of our sample, we also build a mock catalog omitting SPIRE data to evaluate the impact of submm data on constraining the models.
		We present the results from the mock catalogs for $U_{min}$, $\gamma$, $q_{PAH}$, $\log L_{IR}$ and $\log M_{dust}$ in Figure~\ref{mock1} and in Figure~\ref{mock2}.
		For each parameter, we show the case where the photometric coverage is complete (Figure~\ref{mock1}), where there is no 8\microns\ flux density, in the absence of 22-24\microns\ data, the case where there are no 60-70\microns\ data and, finally, in the absence of SPIRE data (Figure~\ref{mock2}).
		For each panel of Figure~\ref{mock1} and Figure~\ref{mock2}, we provide the best linear fit, the Spearman correlation coefficient and the median value and standard deviation of the estimated to true value ratio in Table~\ref{statmock}.
		We separate galaxies having a 24\microns\ measurement from \textit{Spitzer}/MIPS (blue filled points) from galaxies having a 22\microns\ measurement from WISE (red empty points).
		This separation allows to see the impact of the precision of the photometry on the determination of the parameters, especially for $\gamma$ as we will discuss.

		A complete photometric coverage from 12 to 500\microns\ is sufficient to constrain the $U_{min}$ parameter, with a mean ratio between the estimated value to the true one of 1.07$\pm$0.20 (Table~\ref{statmock}).
		The absence of 8\microns\ measurement does not affect the estimation of $U_{min}$.
		The lack of data at 22-24\microns\ yields an under-estimation of the parameter $U_{min}$ of $\approx$14\% whereas the lack of 60-70\microns\ yields an over-estimation of 18\%.
		The worst case is when no SPIRE data is available with an over-estimation of 25\%, showing the importance of submm data to constrain properly $U_{min}$.
		Indeed, as the $U_{min}$ parameter directly probes the position of the IR peak, constraints from both part of the peak are needed to have a good estimation.

		Considering the $\gamma$ parameter, the median ratio between the estimated and true values of the parameters is 1.18.
		However, this median drops to 1.01 when values of $\gamma$ are larger than 0.5\%.
		With our photometric coverage, we tend to over-estimate the low values of $\gamma$. 
		The importance of the 22 and 24\microns\ measurements and the errors associated is clear.
		When all bands are available, the precision of the MIPS 24 photometry gives a good constrain of the estimation of $\gamma$ especially for values above a few percent.
		With a mean error of 13\% on the WISE 22 photometry, there is a good correlation for values of $\gamma$ above a few percents, but the errors on the estimated parameters are very large, indicating a large PDF linked to the estimation of $\gamma$.
		The impact of this 22-24\microns\ range is confirmed in the panel where these two bands are removed, the median value of the estimated to true values ratio is 3.77$\pm$6.43, and the relation is flat.
		When the 22-24\microns\ bands are removed, the estimated gamma derived from the model is rather constant whatever the galaxy we model.
		The absence of 60-70\microns\ or SPIRE data does not have a large impact on the estimation of $\gamma$ but yields to a slightly more dispersed relation.
		
		The 8\microns\ band directly probes the PAH emission: in the absence of the IRAC 8 band, the Spearman correlation coefficient between the estimated $q_{PAH}$ and the true $q_{PAH}$ decreases from 0.92 to 0.88 and the standard deviation of the estimated to true values ratio increases from 0.11 to 0.20.
		Interestingly, the absence of 22-24\microns\ data provides a median under-estimation of the parameter of 23\%, showing the importance of constraining the continuum to estimate the $q_{PAH}$ parameter with our photometric coverage.
		From Figure~\ref{mock1}, we conclude that the presence of the 12-22-24\microns\ data is sufficient to have a relatively good constrain on the fraction of PAH, $q_{PAH}$, provided by the DL07 models.
		For our sample of galaxies, the 8\microns\ measurement is thus not mandatory to study $q_{PAH}$. 
		
		The $L_{IR}$ and the $M_{dust}$ are well constrained in all configurations. 
		Even without the SPIRE bands, the $M_{dust}$ is well constrained despite a slightly larger scatter.
		This constraint on $M_{dust}$ comes from the fact that $\beta$ is fixed in DL07 (to $\approx$2.06) and, as this parameter is provided by the normalization of the models to the observed data, observations at 160\microns\ seem to be sufficient to constrain $M_{dust}$. 
		
		All the parameters are constrained with the combination of IR bands available for this study, except $\gamma$ for which MIPS 24\microns\ measurement is mandatory.
		Therefore, the discussion of $\gamma$ and $<U>$ (as $<U>$ depends on $\gamma$) will be restricted to the galaxies having a 24\microns\ flux density.
		However, we note that this analysis from mocks allows only to characterize our ability to constrain the parameters of DL07 given our photometric coverage.
		
		\subsection{\label{fitresults}Results of the fits}
		
		Table~\ref{result_param} presents the results of the fit for the 146 galaxies: the output parameters from DL07 models, $L_{IR}$ and $M_{dust}$.

			\subsubsection{Assessment of the quality of the fit}
		
			\begin{figure}
				\includegraphics[width=\columnwidth]{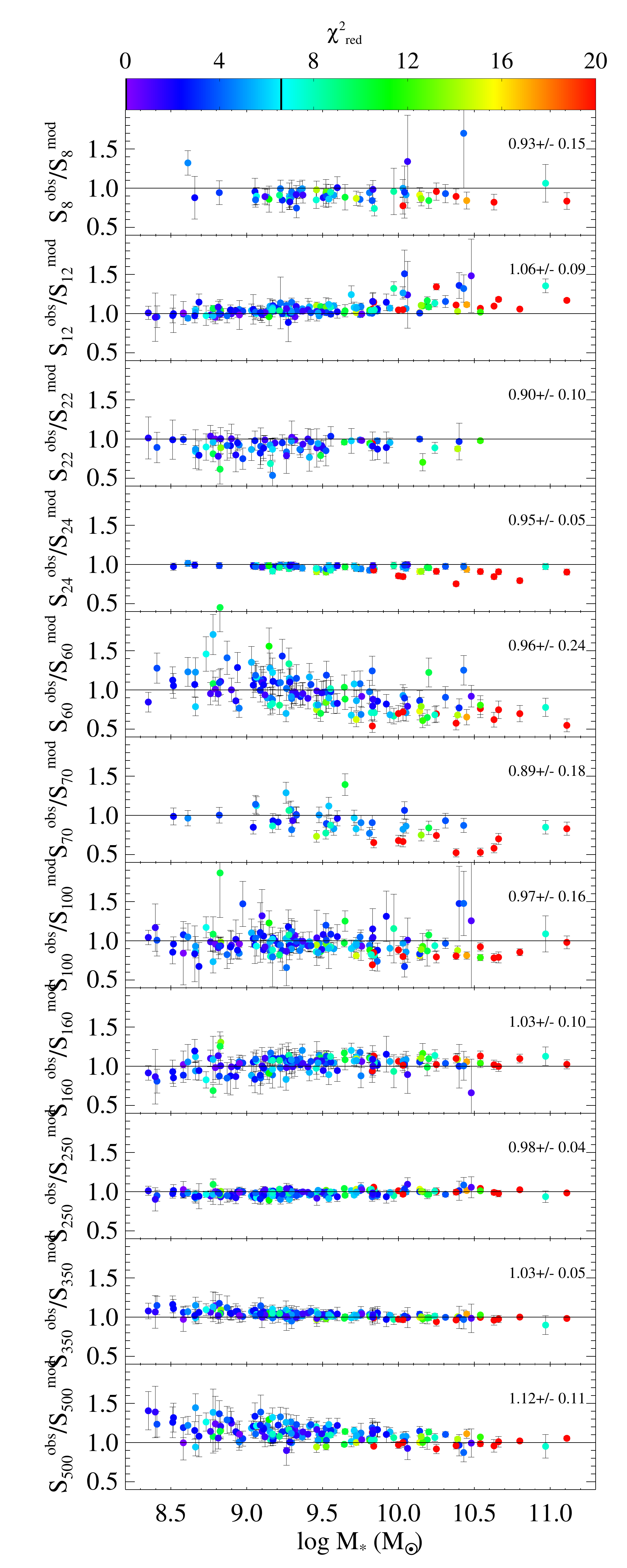}
  			\caption{ \label{modelvsobs} Observed to modeled flux densities ratios for every photometric band (from 8\microns, top panel, to 500\microns, lower panel) versus the stellar mass. Points are color-coded according to the reduced $\chi^2$ associated with the fit. For each wavelength, we indicate the mean value and the standard deviation of the ratios. High mass systems have larger $\chi^2$. DL07 models under-estimate the 500\microns\ observations, especially for low mass systems.}
			\end{figure}

			In order to have a global look at the quality of the fits, we show in Figure~\ref{modelvsobs} the ratio between the observed and the modeled flux densities for each photometric band versus the stellar mass (the calculation of $M_*$ is described in Section~\ref{comp}), as the HRS was selected in K-band which is a proxy for stellar mass.
			For each galaxy, the values predicted by the models are obtained by integrating the best SEDs resulting from the fits into the filters of every band.
			Points are color coded according to the $\chi^2_{red}$ value associated with the fit.
		
			The analysis of Figure~\ref{modelvsobs} shows that:
			\begin{itemize}
		
			\item Fits of galaxies with $\log (M_*/$M$_{\odot})<$10 have a better $\chi^2_{red}$ than higher mass systems for which $\chi^2_{red}$ can reach up to 20.
		
			\item At 8\microns, the models seem to systematically over-predict the observed flux density by $\approx$7\% from low- to high-mass systems.
		
			\item At 12\microns, there is a good agreement between both observed and modeled flux densities up to $\log (M_*/$M$_{\odot})\approx$10.
			For higher mass galaxies, the models under-predict the observed flux densities.
		
			\item There is a systematic over-estimation of the 22-24\microns\ by the models  of 10\% at 22\microns\ and 5\% at 24\microns.
			For the MIPS 24\microns, this over-estimation increases with the stellar mass and can reach up to 20\%.
			Despite this trend, with a mean ratio of 0.95 and a small dispersion 0.05, there is a good agreement between the models and the observations at this wavelength.
		
			\item The 60\microns\ ratio is very dispersed, and the mean value of 0.96 is not representative as there is a clear trend with the stellar mass.
			Models under-predict the 60\microns\ observations of low mass systems and over-predict them for high mass objects.
			However, the mean photometric error of 15\% is one of the largest of all bands and is not a strong constraint for the fit. 
			The 70\microns\ observed to modeled flux density ratios follow the same trend with the stellar mass than the 60\microns, but with a smaller dispersion ($\sigma$=0.18).
		
			\item At 100 and 160\microns, despite a large dispersion, there is a relatively good agreement between observations and models with a slight over-prediction at 100\microns\ and under-prediction at 160\microns\ of $\approx$3\%.
		
			\item The 250 and 350\microns\ SPIRE bands are in average well fitted by the models, even if the 250\microns\ is slightly over-estimated for the whole sample.
			A small trend is seen at 350\microns, the flux densities from the models tend to be lower than the observations for low mass galaxies. 
		
			\item At 500\microns, this trend becomes stronger.
			The 500\microns\ observations seem to be under-predicted by the models, especially for low-mass systems.	
			Indeed, for galaxies with $\log (M_*/$M$_{\odot})\approx$10, there is a clear under-prediction of the 500\microns\ data that increases when the stellar mass decreases up to $\approx$40\%.
			\end{itemize}
		
			A submm excess, such as the one observed at 500\microns, was already noticed by previous works in low-metallicity systems \citep[e.g.][]{Reach95,Galliano03,Galliano05,Galliano11,Bot10,Gordon10,Boselli10a,Galametz11,Boselli12}.
			Different hypothesis have been proposed to explain this excess.
			A very cold dust component was proposed by \cite{Galliano03,Galliano05}, showing that it would need to lie in a small number of dense parsec scale clumps, but \cite{Galliano11} showed that this hypothesis  was not verified in the LMC, where they had the spatial resolution to test it.
			\cite{Meny07} proposed a solid state based temperature dependent emissivity increase at long wavelength, in amorphous materials.
			Finally, \cite{DraineHensley12} argued that this excess could be due to ferromagnetic free flying small grains or ferromagnetic inclusions in normal grains.
			Changing the FIR $\beta$ slope can make the excess diminish but it is not sufficient \citep{Galliano11}.
			\cite{Boselli12} noticed that a modified black body with an emissivity index of $\beta$=1.5 better represents the SPIRE data of the HRS galaxies than the DL07 models which submm slope can be approximated by $\beta \approx$2.
			However, they outlined that $\beta$=2 is possible for metal-rich high-mass galaxies, confirming what we observe in Figure~\ref{modelvsobs}.
			\cite{Remy-Ruyer13} showed with a sample of galaxies with metallicities ranging from 0.03 to 1\,Z$_{\odot}$ that $\beta$ obtained from modified black body fit shows a large spread from 0.5 to 2.5.
					
			These disagreements between the observations and the models are difficult to interpret as they can have different origins.
			On one hand, the way that errors on the photometry are computed determine the relative weights of the different bands, and thus play an important role to constrain the models, the case of the 22 and 24\microns\ being one illustration.
			On the other hand, the trend between the stellar mass and the $\chi^2_{red}$ also suggests that the models do not reproduce very well the observations of high-mass galaxies, and the 500\microns\ observations of low-mass systems are not reproduced by the DL07 models.

\tiny
\onecolumn{
\begin{longtable}{c c c c c c c c c c c c }
\caption{\label{result_param} Estimated parameters.} \\
\hline
\hline
HRS  & CGCG   &  VCC   & UGC  &  NGC  &  IC   & Type  & $U_{min}$ &  $q_{PAH}$ (\%) & $\gamma$ (\%) & $\log (L_{IR}/\rm{L_{\odot}})$ &  $\log (M_{dust}/\rm{M_{\odot}})$  \\
\hline
 \endfirsthead
\caption{continued.}\\  
\hline
\hline
HRS  & CGCG   &  VCC   & UGC  &  NGC  &  IC   & Type  & $U_{min}$ &  $q_{PAH}$ (\%) & $\gamma$ (\%) & $\log (L_{IR}/\rm{L_{\odot}})$ &  $\log (M_{dust}/\rm{M_{\odot}})$  \\
\hline
\endhead
\hline
\endfoot
       2       &    124004       &         -       &      5588       &         -       &         -       &        Sb       &            5.80   $\pm$            1.45       &            4.53   $\pm$            0.18       &            0.51   $\pm$            0.39       &            9.20   $\pm$            0.02       &            6.25   $\pm$            0.08       \\
       8       &    154020       &         -       &      5685       &      3254       &         -       &       Sbc       &            0.38   $\pm$            0.10       &            4.46   $\pm$            0.30       &            0.40   $\pm$            0.29       &            9.25   $\pm$            0.04       &            7.50   $\pm$            0.09       \\
      10       &    183028       &         -      &      5738       &         -      &         -       &        Sb       &            2.60   $\pm$            1.33       &            4.39   $\pm$            0.35       &            1.39   $\pm$            1.42       &            8.94   $\pm$            0.04       &            6.29   $\pm$            0.15       \\
      11       &    124038       &         -       &      5742       &      3287       &         -       &        Sd       &            2.09   $\pm$            0.40       &            4.47   $\pm$            0.25       &            0.66   $\pm$            0.50       &            9.49   $\pm$            0.02       &            6.97   $\pm$            0.05       \\
      12       &    124041       &         -       &         -       &         -       &         -       &       BCD       &            3.52   $\pm$            1.34       &            3.10   $\pm$            0.97       &            1.00   $\pm$            1.13       &            8.52   $\pm$            0.05       &            5.77   $\pm$            0.10       \\
\hline
\end{longtable}
}
\twocolumn
\normalsize

			\subsubsection{Statistics of the output parameters}
				
			\begin{table}
				\centering
				\begin{threeparttable}[b]
				\caption{Statistics from the fitting of the 146 normal LTGs.}
				\begin{tabular}{l c c c c }
		  		\hline\hline
				\multicolumn{1}{c}{} 	& \multicolumn{4}{c}{Best fit from $\chi^2$}\\ 
				\hline 
				\multicolumn{1}{c}{} 	& Median & $\sigma$& Min & Max\\ 
				\hline		
				$\chi^2_{red}$		& 2.25 & 5.33 & 0.10 & 42.72\\			
				$U_{min}$			& 2.00 & 1.70 & 0.20 & 8.00 \\
				$q_{PAH}$ (\%)		& 4.58 & 0.85 & 0.47 & 4.58\\
		  		$\gamma$ (\%)		& 0.76 & 1.73 & 0.1 & 12.3 \\
				\hline
				$\log (L_{IR}/$L$_{\odot})$		& 9.48 & 0.54& 8.21 & 10.96 \\
				$\log (M_{dust}/$M$_{\odot})$		& 7.07 & 0.50 & 5.77 & 8.42 \\
		  		\hline
				\hline
				\multicolumn{1}{c}{} 	&  \multicolumn{4}{c}{Estimation from the PDF} \\
				\hline  
				\multicolumn{1}{c}{} 	&  Median & $\sigma$& Min & Max\\ 
				\hline		
				$U_{min}$			& 1.96$\pm$0.47 &1.76 &0.26  &8.58 \\
				$q_{PAH}$ (\%)		& 4.16$\pm$0.42 &0.76 &1.22  &4.58\\
		  		$\gamma$ (\%)		& 0.75$\pm$0.5 &2.03 &0.23  &14.7 \\
				\hline
				$\log (L_{IR}/$L$_{\odot})$	& 9.47$\pm$0.03 &0.53 &8.24  &10.94 \\
				$\log (M_{dust}/$M$_{\odot})$	& 7.02$\pm$0.08 &0.48 &5.77  &8.38 \\
		  		\hline
				\label{stat}
				\end{tabular}
				\end{threeparttable}
			\end{table}

			The statistics of the fits and of the derived parameters are presented in Table~\ref{stat}.
			The median $\chi^2_{red}$ is 2.25.
			The median estimated value of $U_{min}$ is 1.96$\pm$0.47.
			If we consider the relation obtained by \cite{Aniano12}, this median $U_{min}$ corresponds to a median temperature of the dust of 22\,K.
			We find a median estimated $q_{PAH}$ of 4.16$\pm$0.42\%.
			We obtain a median $\gamma$ of 0.75$\pm$0.5\% when considering only the 24\microns\ sample. 
			Finally, this late-type non-deficient sample is characterized by a median $\log (L_{IR}/$L$_{\odot})$ of 9.47$\pm$0.03 and $\log (M_{dust}/$M$_{\odot})$ of 7.02$\pm$0.08.
			We compare the dust masses we derive with those obtained for the same galaxies by \cite{Cortese12a} using only SPIRE bands and find a good agreement.
			The median ratio between the $M_{dust}$ of \cite{Cortese12a} and ours is 1.02, with a standard deviation of 0.02.
		
\begin{figure}
				\includegraphics[width=9cm]{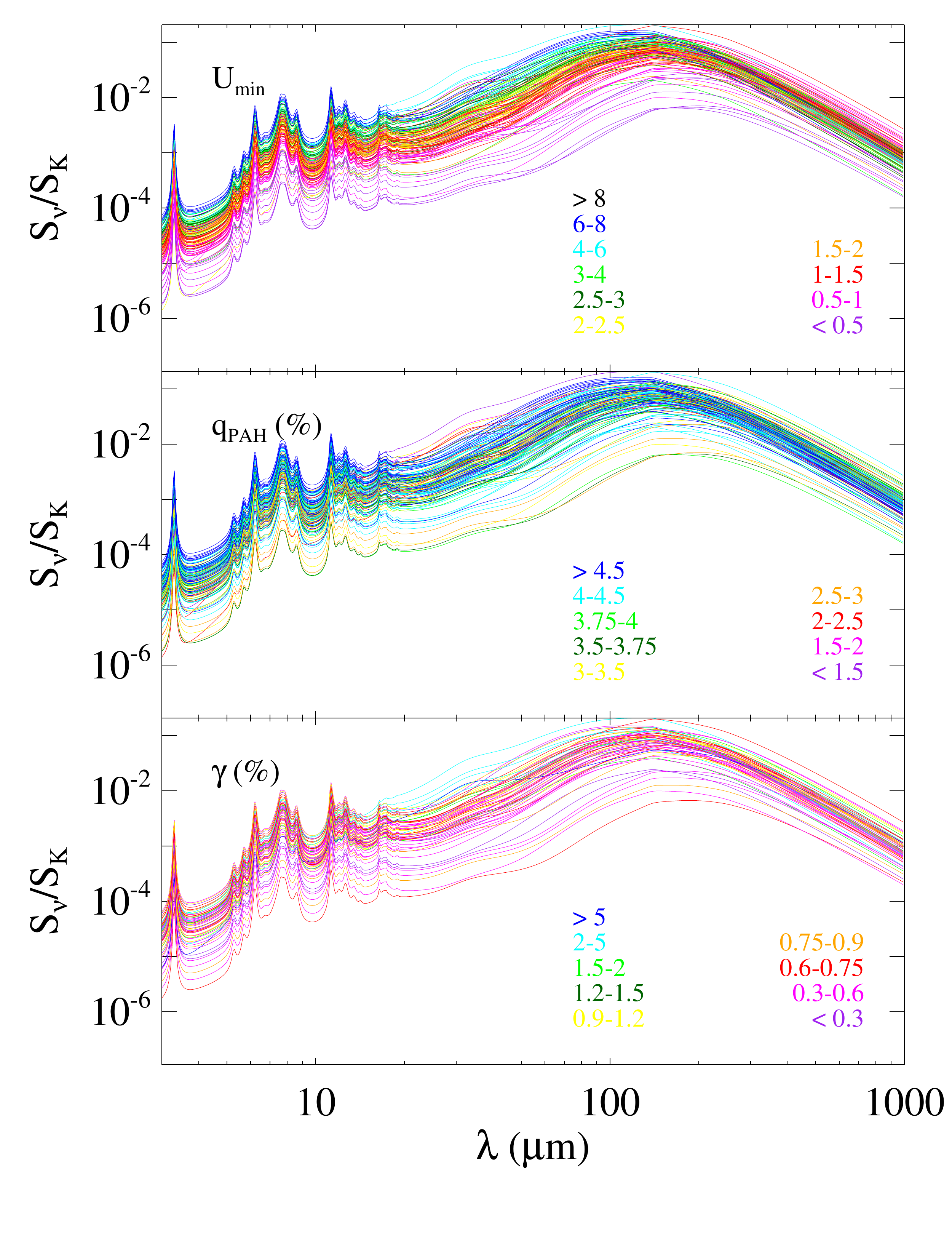}
  				\caption{ \label{sedfit1} Best fit SEDs normalized to the observed K-band flux densities and color coded according to the estimated value of the DL07 output parameter (top pannel: $\gamma$, middle panel: $q_{PAH}$ and lower panel: $U_{min}$).}
			\end{figure}

\begin{figure}
				\includegraphics[width=9cm]{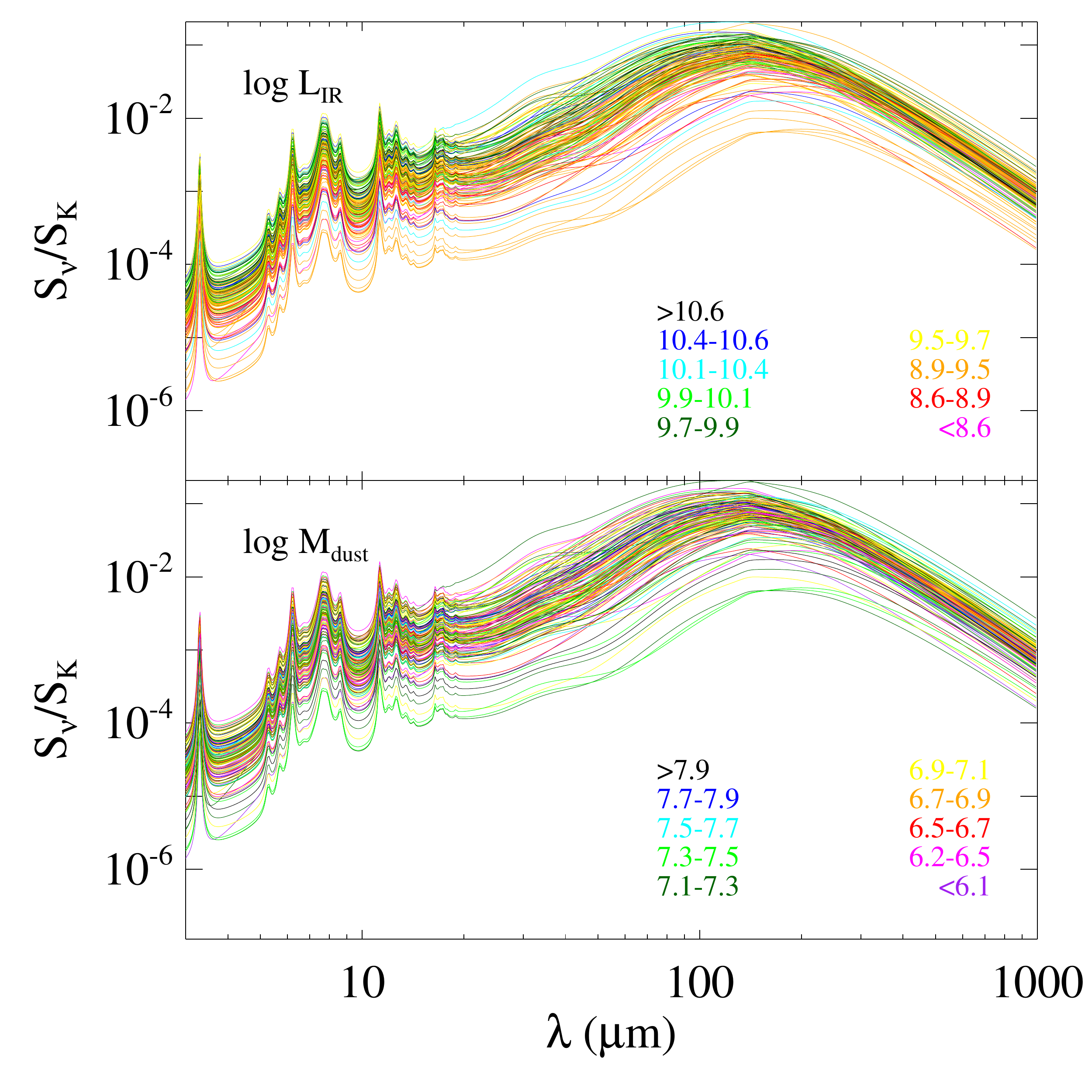}
  				\caption{ \label{sedfit2} Best fit SEDs normalized to the observed K-band flux densities and color coded according to the value of the resulting $\log L_{IR}$ (top panel) and $\log M_{dust}$ (lower panel). }
			\end{figure}

			The SEDs obtained from the minimum $\chi^2_{red}$ fit for each galaxy are shown in Figure~\ref{sedfit1} and Figure~\ref{sedfit2}.
			They are normalized to the observed 2MASS K-band flux densities as the galaxies of our sample were selected in this band \citep[see][ for the description of the K band flux densities]{Boselli10a}.
			In each panel, we show the best fitted SEDs, color-coded according to the estimated values of the three output parameters from the models ($U_{min}$, $q_{PAH}$ and $\gamma$), plus the $L_{IR}$ and the $M_{dust}$ directly derived from the fits.
		
			We notice a color gradient with $U_{min}$, indicating a relation with the shape of the SED.
			By definition, $U_{min}$ controls the minimum and dominant equilibrium dust temperature, therefore controlling the wavelength peak of the SED, as we can see on the top panel of Figure~\ref{sedfit1}.
			Middle panel of Figure~\ref{sedfit1} shows that $q_{PAH}$ parameter impacts the shape of the SED handling the intensity of the PAH bands.
			However, our sample is dominated by galaxies with $q_{PAH}>$4\%.
			This concerns 91 out of 146 galaxies, implying that there is a need for models with a larger range of PAH.
			Furthermore, the most massive galaxies, with $\log (M_*/$M$_{\odot})>$10, have a PAH fraction $>$4\% and the larger $\chi^2$ values.
			This suggests that the lesser quality of the fit of the most massive galaxies is due to the small range of $q_{PAH}$ values available.
			An increase in $\gamma$ translates at a bump in the 25-60\microns\ range, as shown in the bottom panel of Figure~\ref{sedfit1}.
			The lack of observational data in this range can explain the difficulty to constrain $\gamma$, already outlined in Section~\ref{mocks}.
			
			The fitting procedure also provide us with the $L_{IR}$ and the $M_{dust}$ (Figure~\ref{sedfit2}).
			A trend between the $L_{IR}$ and the shape of the SED is visible.
			Late-type galaxies are more gas rich, and thus have a higher dust content, therefore they form more stars.
			The tight link between the $SFR$ and the $L_{IR}$ yields to the trend observed in the top panel of Figure~\ref{sedfit2}.
			No particular relation is found between $M_{dust}$ and $S_{\nu}/S_K$.
		
			\subsubsection{Comparison with the literature}
		
			This sample and the SINGS/KINGFISH \citep{Kennicutt03,Kennicutt12} sample have 4 galaxies in common: M\,99, M\,100, NGC\,4536 and NGC\,4725.	
			We compare, in Table~\ref{kingfishhrs}, the results from the fits from our procedure with those found by \cite{Draine07} and \cite{Dale12} fitting DL07 models.
			For M\,99, M\,100 and NGC\,4725, \cite{Draine07} did not have any submm constraint on the SED, except for NGC\,4536 for which they had a SCUBA 850\microns\ flux density.
			\cite{Dale12} benefited from a larger set of models yielding to a larger parameter space to be explored with $0.01 < U_{min} < 30$, $3 < \log U_{max} < 8$ and $0\% < q_{PAH} < 12\%$.
			Despite the differences with the photometric coverage for SINGS and with the parameter ranges for KINGFISH, we find results in agreement.

			\begin{table}
				\begin{center}
				\caption{Comparison of the outputs parameters from DL07 determined in this work, by \cite{Dale12} (KINGFISH) and by \cite{Draine07} (SINGS) for 4 galaxies in common.}
				\begin{tabular}{l l c c c}
	  			\hline\hline
				Galaxy &Parameter & HRS & KINGFISH & SINGS\\ 
				\hline		
				M\,99		&$U_{min}$				&4.49$\pm$0.50& 3.0	 &1.5\\			
							&$q_{PAH}$ (\%)			&4.58$\pm$0.00\tablefootmark{a} & 6.9 &4.5\\
							&$\gamma$	(\%)			&0.73$\pm$0.30 & 0.8 &0.96\\
	  						&$\log (L_{IR}/$L$_{\odot})$		 	&10.83$\pm$0.01 &10.6 &10.91\\
							&$\log (M_{dust}/$M$_{\odot})$			&7.97$\pm$0.04 &7.95 &8.55\\
				\hline
				M\,100		&$U_{min}$				&2.47$\pm$0.35& 2.0 &1.2\\			
							&$q_{PAH}$	(\%)			&4.58$\pm$0.00\tablefootmark{a} & 5.9 &4.2\\
							&$\gamma$	(\%)			&0.65$\pm$0.28 & 0.9 &0.85\\
	  						&$\log (L_{IR}/$L$_{\odot})$		 	&10.72$\pm$0.01 &10.5 &10.82\\
							&$\log (M_{dust}/$M$_{\odot})$			&8.12$\pm$0.06 &8.07 &8.57\\
				\hline			
				NGC\,4536	&$U_{min}$				&2.58$\pm$0.48 &3.0 & 5.0\\			
							&$q_{PAH}$	(\%)			&4.16$\pm$0.42 &4.3 & 3.5\\
							&$\gamma$	(\%)			&3.29$\pm$0.87 &2.7 & 3.31\\
	  						&$\log (L_{IR}/$L$_{\odot})$		 	&10.45$\pm$0.02 &10.3 &10.79\\
							&$\log (M_{dust}/$M$_{\odot})$		&7.72$\pm$0.05 &7.61 &7.82\\
				\hline
				NGC\,4725	&$U_{min}$				&0.52$\pm$0.15 &0.6 &0.7\\			
							&$q_{PAH}$ (\%)			&4.43$\pm$0.33 &6.6 &4.5 \\
							&$\gamma$ (\%)			&0.39$\pm$0.25 &0.4 &0.16\\
	  						&$\log (L_{IR}/$L$_{\odot})$		 	&10.12$\pm$0.02 &9.9  &10.18\\
							&$\log (M_{dust}/$M$_{\odot})$		&8.22$\pm$0.12 &7.98 &8.20 \\
				\hline			
				\label{kingfishhrs}
				\end{tabular}
				\end{center}
				\tablefootmark{a}{For M\,99 and M\,100, the PDF of $q_{PAH}$ is very narrow and yields to an error of 0.0. This emphasize the need for higher fractions of PAH for these galaxies.}
			\end{table}

	\section{\label{comp}Comparison with physical parameters}
	
	We study here the relations between the output parameters of the models and various physical variables available for our sample.
	In a phenomenological model independent approach, the relations between these physical properties and the FIR colors of the gas rich late-type galaxies have been analyzed in \cite{Boselli12} where a complete description of these variables can be found.
	Here we provide a very brief description.
	The number of galaxies of our sample for which ancillary data are available are given in Table~\ref{prop}.
	
	\begin{table}
		\centering
		\caption{Completeness of the integrated properties for the final sample of 146 galaxies.}
		\begin{tabular}{l c }
	  	\hline\hline
		Properties & Number of galaxies\\ 
		\hline		
		$M_{*}$				&146 \\			
		SFR					& 141\\
		$b$					& 141\\
	  	$\Sigma(H\alpha)$ 	& 129\\
		$\mu_e(H)$			& 146\\
		$12+\log(O/H)$		& 124\\
		$A(FUV)$			& 117\\
		$L_{IR}$				& 146\\
		$M_{dust}$			& 146\\
		\hline
		\label{prop}
		\end{tabular}
	\end{table}
	
		\begin{figure*}
		\includegraphics[width=18cm]{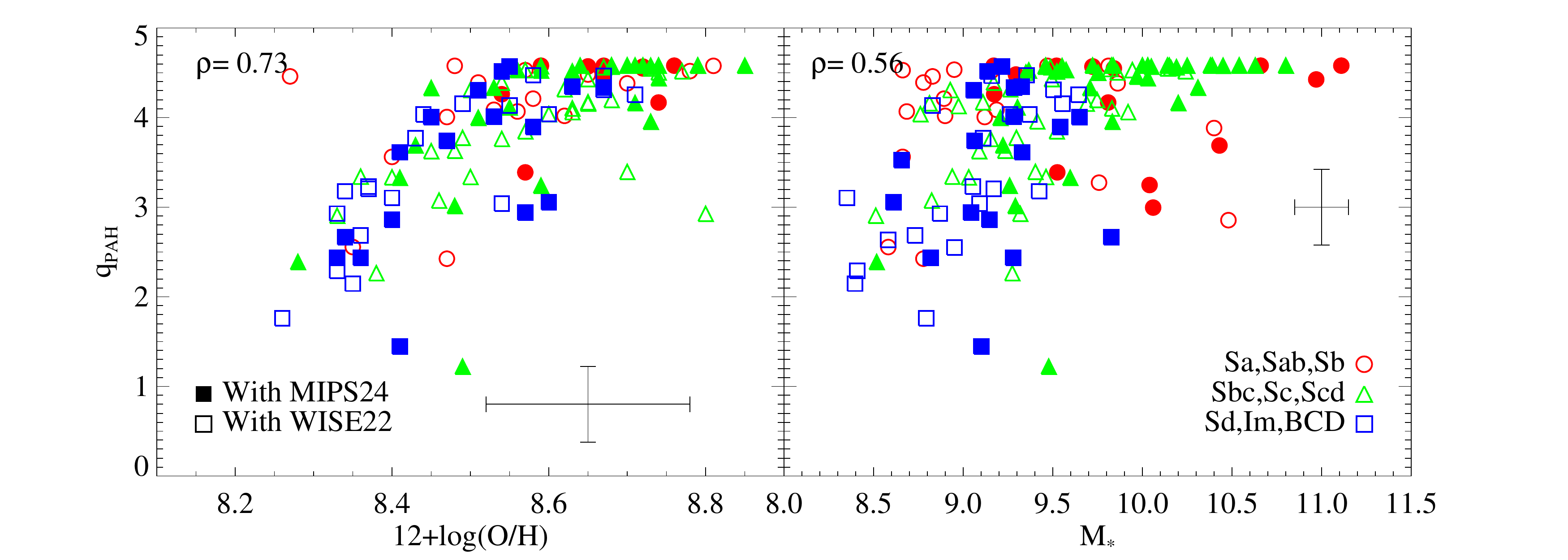}
  		\caption{ \label{metalvsqpah} Relation between the fraction of PAH and the metallicity (left panel) and between the fraction of PAH and the stellar mass (right panel). Galaxies are color-coded according to their morphological type. Red: Sa-Sb, green: Sbc-Scd, and blue: Sd-BCD. Filled symbols are for galaxies with a MIPS 24\microns\ observations and thus a good constraint on $\gamma$. Empty symbols are for galaxies with a WISE 22\microns\ observations. The Spearman correlation coefficients, $\rho$, are provided. The cross gives the typical error bar on the data.}
	\end{figure*}
		
	\begin{figure*}
		\includegraphics[width=18cm]{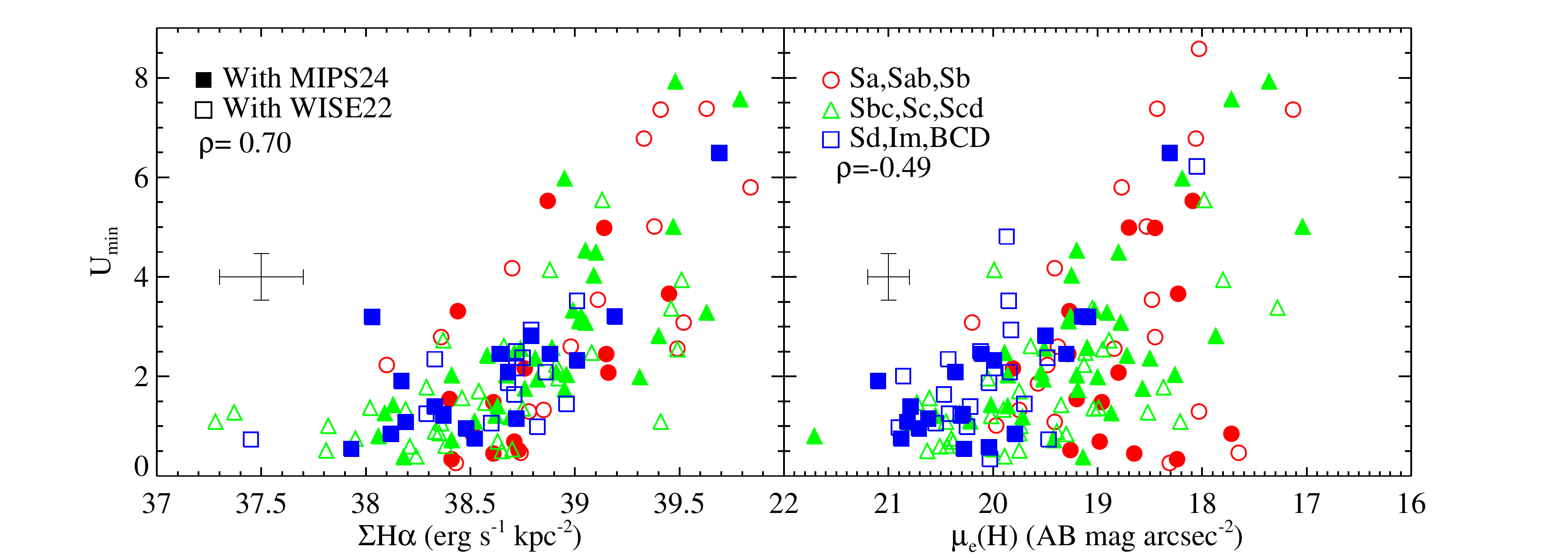}
  		\caption{ \label{uminvssighamueh}  Relation between $U_{min}$, tracer of the minimum intensity of the ISRF,the H$\alpha$ and H-band surface brightness (left panel and right panel, respectively). Galaxies are color-coded according their morphological type. Red: Sa-Sb, green: Sbc-Scd, and blue: Sd-BCD. Filled symbols are for galaxies with a MIPS 24\microns\ observations and thus a good constraint on $\gamma$. Empty symbols are for galaxies with a WISE 22\microns\ observations. The Spearman correlation coefficient, $\rho$, is provided. The cross gives the typical error bar on the data.}
	\end{figure*}
		
	Stellar masses ($M_{*}$) are estimated from H-band data, and from mass-luminosity relations, as determined by the chemo-spectro-photometric galaxy evolution models of \cite{BoissierPrantzos00}, using the relation given in \cite{Boselli09}. 	
	The effective surface brightness in H-band, $\mu_e(H)$, provides a measurement of the intensity of the ISRF produced by the old stellar population. 
	The H$\alpha$ surface brightness is a tracer of the intensity of the ionizing UV emission and provides us with information on the present star formation activity \citep{Boselli09}.
	SFRs are calculated using the standard calibration of \cite{Kennicutt98} to convert H$\alpha$ and FUV luminosities corrected from dust attenuation using the Balmer decrement \citep{Boselli09} and the corrections of \cite{Cortese08}, respectively.
	The final SFR corresponds to the mean value of the SFR determined from the H$\alpha$ data and the FUV data \citep{Boselli09}.
	
	The birthrate parameter $b$ \citep{Scalo86,Kennicutt98}, defined as the ratio between the present star formation activity and the star formation activity averaged on the galaxy entire life, is defined by \cite{Boselli01}:
	\begin{equation}
		b=\frac{SFR}{<SFR>}=\frac{SFR t_0 (1-R)}{M_{*}},
	\end{equation}
	with $t_0$ the age of the galaxy, $M_{*}$ its stellar mass, and $R$ (assumed to be 0.3) the fraction of re-injected gas into the ISM due to stellar winds.
	The SFR being linked to the UV or H$\alpha$ fluxes and $M_*$ to the NIR fluxes, $b$ is thus tightly linked to the hardness of the UV radiation field.
	Moreover, the specific star formation rate \citep[sSFR,][]{Brinchmann04} is widely used and can be linked to the birthrate parameter as:
	\begin{equation}
		sSFR = \frac{SFR}{M_*}=\frac{b}{t_0(1-R)}.
	\end{equation}	

	\cite{Hughes13} measured the metallicity of the galaxies of our sample with integrated spectroscopy, where different calibrations are used to derive $12+\log(O/H)$, depending on the availability of the main emission lines.
	$A(FUV)$, in magnitude, is the attenuation of the non-ionizing UV radiation, defined as the ratio between the FIR and the FUV (at 1539\AA) flux densities, following the recommendations of \cite{Cortese08}.

	The outputs of DL07 models are compared to these integrated properties.
	We decompose our sample in three groups according to the morphological types of the galaxies: Sa, Sab and Sb, Sbc Sc and Scd, and Sd, Im and BCD\footnote{We follow the groups and symbols used in \cite{Boselli12} to allow a better comparison between the two works.}.
	Galaxies with a MIPS24 measurement are represented by filled symbols whereas galaxies with WISE22 are represented by empty symbols.
	All the relations are presented in Figure~\ref{totparam}, and discussed in Appendix~\ref{apcompparam}.
	For the total sample, we give the Spearman correlation coefficient of the relations.
	With a number of objects larger than 100, a correlation is expected to be real with a Spearman correlation coefficient larger than 0.40.
	We outline here the most interesting correlations between the properties and the DL07 output parameters.
	
	One of the strongest trend is found between the metallicity and $q_{PAH}$ ($\rho$=0.73, Figure~\ref{metalvsqpah}), 
	Metal-rich galaxies, with 12+$\log(O/H)$ have a high fraction of PAH.
	PAH are destroyed in low metallicity environment by the UV radiation field which propagates more easily due to the lower dust content.
	This was demonstrated in \cite{Galliano03}, \cite{Boselli04}, \cite{Madden06}, \cite{Engelbracht08} and \cite{Gordon08} for galaxies with lower metallicities. 
	We should note that, in this work, we observe this relation for metal-rich galaxies with 8.3$<$12+$\log$(O/H)$<$8.85.
	We also confirm on a larger sample the results of \cite{Engelbracht05} who found a weakening of the PAH features at low metallicity from 8 to 24\microns\ flux density ratios, and \cite{Smith07} who found that the strength of the PAH bands is directly linked to the metallicity from the \textit{Spitzer}/IRS spectrum of 59 nearby galaxies.	
	However, \cite{Galliano08} suggested that the absence of PAH could be due to a delayed injection of carbon dust by AGB stars, and \cite{Sandstrom12} proposed that the PAH were formed in molecular clouds, which have a lower filling factors in low-metallicity environments.
	A good relation is found between the stellar mass and $q_{PAH}$ ($\rho$=0.56).
	Indeed, massive galaxies are also more metal-rich \citep{Tremonti04}.
	
	A good correlation is found between the $U_{min}$ parameter, which is the intensity of the diffuse ISRF \citep{Draine07,Aniano12}, and the H$\alpha$ surface brightness ($\rho=$0.70), and a moderate anti-correlation between $U_{min}$ and the H-band surface brightness in AB\,mag\,arcsec$^{-2}$ ($\rho=$-0.49) as shown in Figure~\ref{uminvssighamueh}.
	
	A similar behavior was outlined by \cite{Boselli12} where they found relations between FIR colors sensitive to the IR emission peak ($S_{60}/S_{250}$, $S_{60}/S_{100}$ and $S_{100}/S_{500}$) and $\Sigma H\alpha$ and $\mu_e(H)$.
	This is consistent with what we observe in this work as the $U_{min}$ parameter regulates the position of the IR emission peak (see Figure~\ref{sedfit1}).
	The stars of the diffuse component emit the bulk of their radiation in NIR, which can be probed by the H-band surface brightness.
	It is thus expected to find a relation between $U_{min}$ and $\mu_e(H)$.
	This means that the diffuse dust is heated by the old stellar component.
	Furthermore, a good relation is also found with $\Sigma H\alpha$ which is the ionizing surface brightness due to the young stars.
	The diffuse dust component seems to be also heated by the young stars component in star forming regions.
	With a Spearman coefficient of -0.44, there is not a strong correlation between $\mu_e(H)$ and $\Sigma H\alpha$ that could have originate the relations with $U_{min}$.
	This result from integrated galaxies seems incompatible with works based on the analysis of resolved galaxies \citep{Bendo10,Bendo12b,Boquien11}.
	These studies showed that at wavelengths shorter than 160\microns, most of the dust is heated by massive stars whereas at wavelengths longer than 250\microns, the dust is primarily heated by the evolved stellar populations.
	This discrepancy can be explained by the fact that the brightest regions of the galaxies dominate the emission measured with integrated flux densities.
	For late-type galaxies, in IR, the brightest regions are heated by star formation and thus are linked to the H$\alpha$ emission, explaining the correlation that we observe.
	Therefore, enhanced star formation might increase $U_{min}$, in which case the interpretation of $U_{min}$ may be that it no longer traces only the diffuse ISRF from evolved stars but just the lowest energy radiation field within the galaxies for integrated studies.
	
\section{\label{temp}Infrared templates}

	\begin{figure}
			\includegraphics[width=6cm,angle=90]{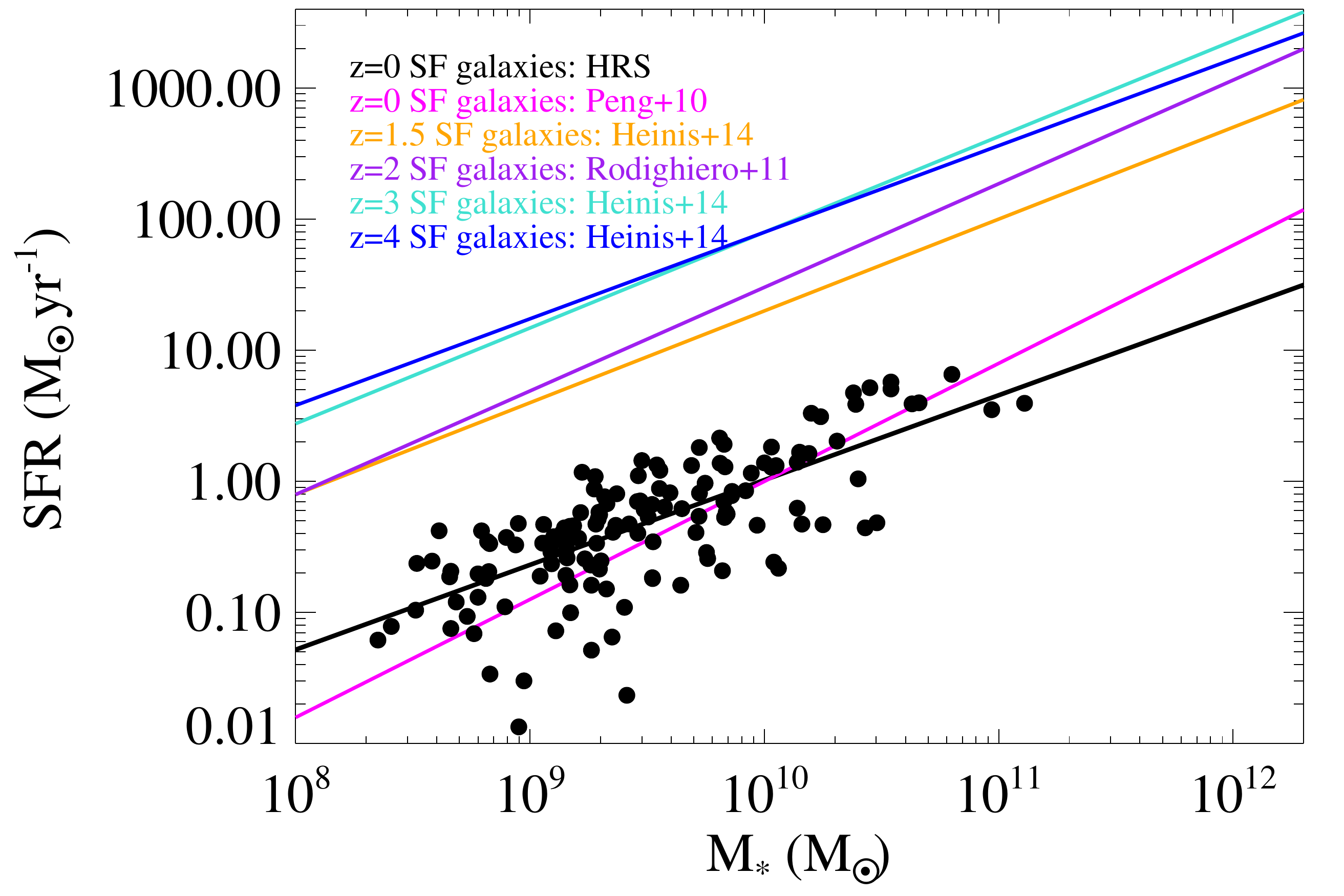}
  			\caption{ \label{MS} HRS galaxies placed on the SFR-$M_*$ plot (black dots). The best linear fit is shown as a black solid line. For comparison, other MS fits estimated at different redshifts are shown. The HRS galaxies are consistent with the relation of \cite{Peng10}, however, it seems that HRS low mass systems have higher SFRs.}
	\end{figure}
	
	Thanks to the wealth of photometric data and physical properties available for the HRS galaxies, the gas-rich galaxy sample studied in this work is ideal to build new, well-constrained, IR templates.
	The majority of star-forming galaxies are known to follow a SFR-$M_*$ correlation, called the main sequence (MS), and galaxies that lie above this sequence are experiencing a starburst event \citep{Elbaz11}.
	Several studies have put constraints on the MS at different redshifts \cite[e.g.,][]{Peng10,Rodighiero11,Heinis14}.
	Thanks to its design, volume-limit and completeness in the K-band, the HRS is well suited to provide constraints on the local MS as the results will be relatively free of distance-related biases that could appear in flux-limited sample of galaxies.
	Figure~\ref{MS} puts the late-type star-forming galaxies of our sample in a SFR-$M_*$ plot, along with several MS estimations.
	The SFRs are calculated assuming a Salpeter IMF and the stellar masses assuming an IMF of Kroupa, we thus corrected the SFRs by a factor of $-0.17$\,dex, following the results of \cite{Brinchmann04} and \cite{Buat14}.
	A direct comparison can be done only with the sample of \cite{Peng10} since it is the only one based on local galaxies.
	The MS local estimation of \cite{Peng10} was built from a subsample of SDSS galaxies with 0.02$< z< $0.085.
	Their selection is only complete at $z$=0.085 above a stellar mass of $\approx$10$^{10.4}$\,M$_{\odot}$.
	There is a relatively good agreement with the MS local estimation of \cite{Peng10}.
	The observed difference in the slope may come from the different ranges of stellar masses that our sample and the sample of \cite{Peng10} cover.
	The best linear fit determined from our galaxies is:
	\begin{equation}
		\log SFR = 0.65\times \log M_* -6.46.
	\end{equation}

	\subsection{Construction of the templates}

	\begin{figure}
		\includegraphics[width=6cm,angle=90]{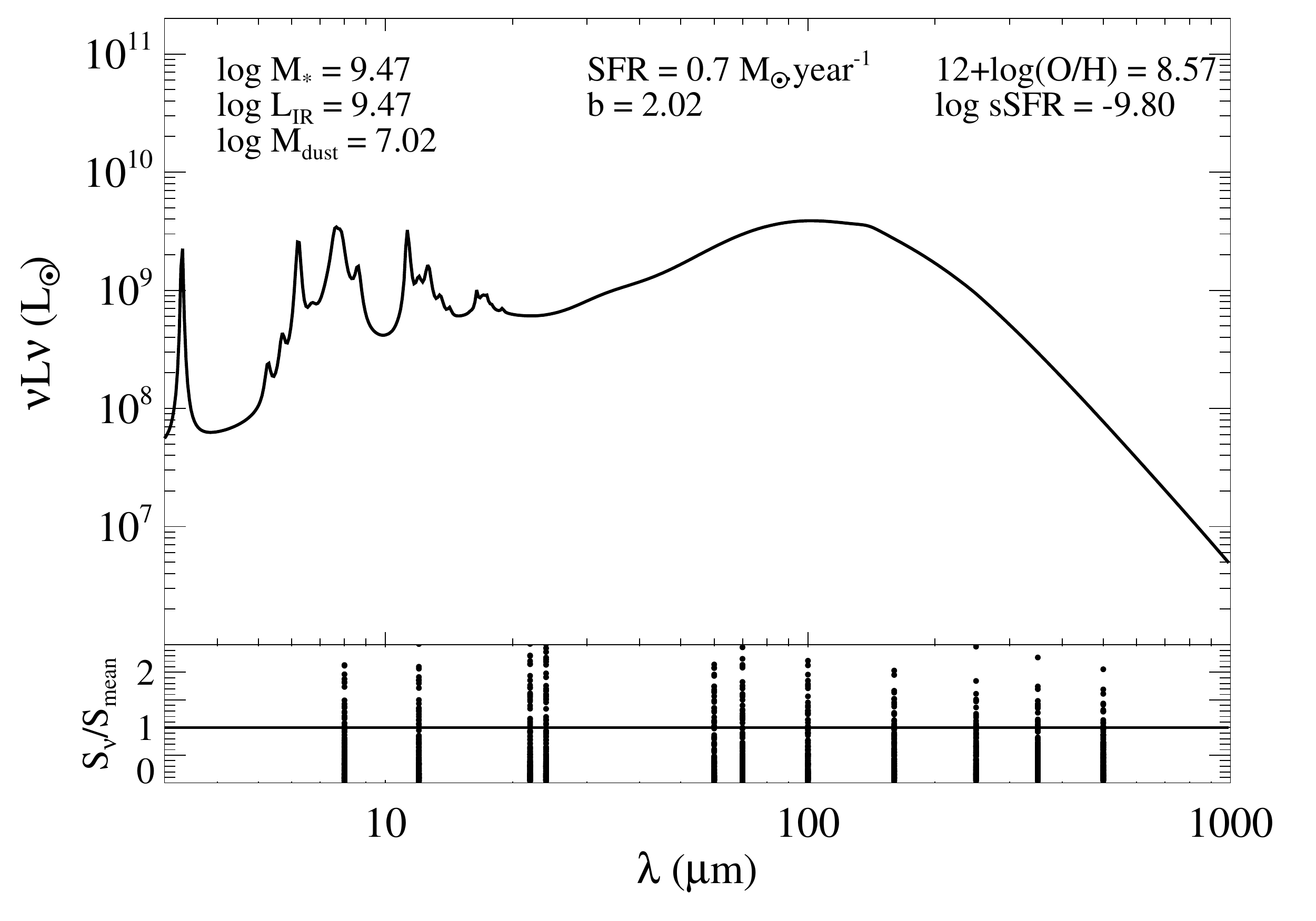}
  		\caption{ \label{template_mean}Top panel: Mean SED of $z$=0 normal star forming galaxies. The mean values of the physical parameters associated to the sample are indicated. Bottom panel: Dispersion of the models used to compute the mean SED, in all of the bands of our photometric coverage.}
	\end{figure}
	
	We computed the mean SED of the whole sample of late-type non deficient galaxies from this work in order to provide the typical $z$=0 SED of nearby normal galaxies (Figure~\ref{template_mean}).
	To do so, we averaged all the best fit SEDs of the gas-rich subsample studied in the previous Sections.
	We provide the mean values of all of the physical properties that are associated to the sample from which this mean SED originates (Figure~\ref{template_mean}).

	\begin{figure*}
		\includegraphics[width=19cm]{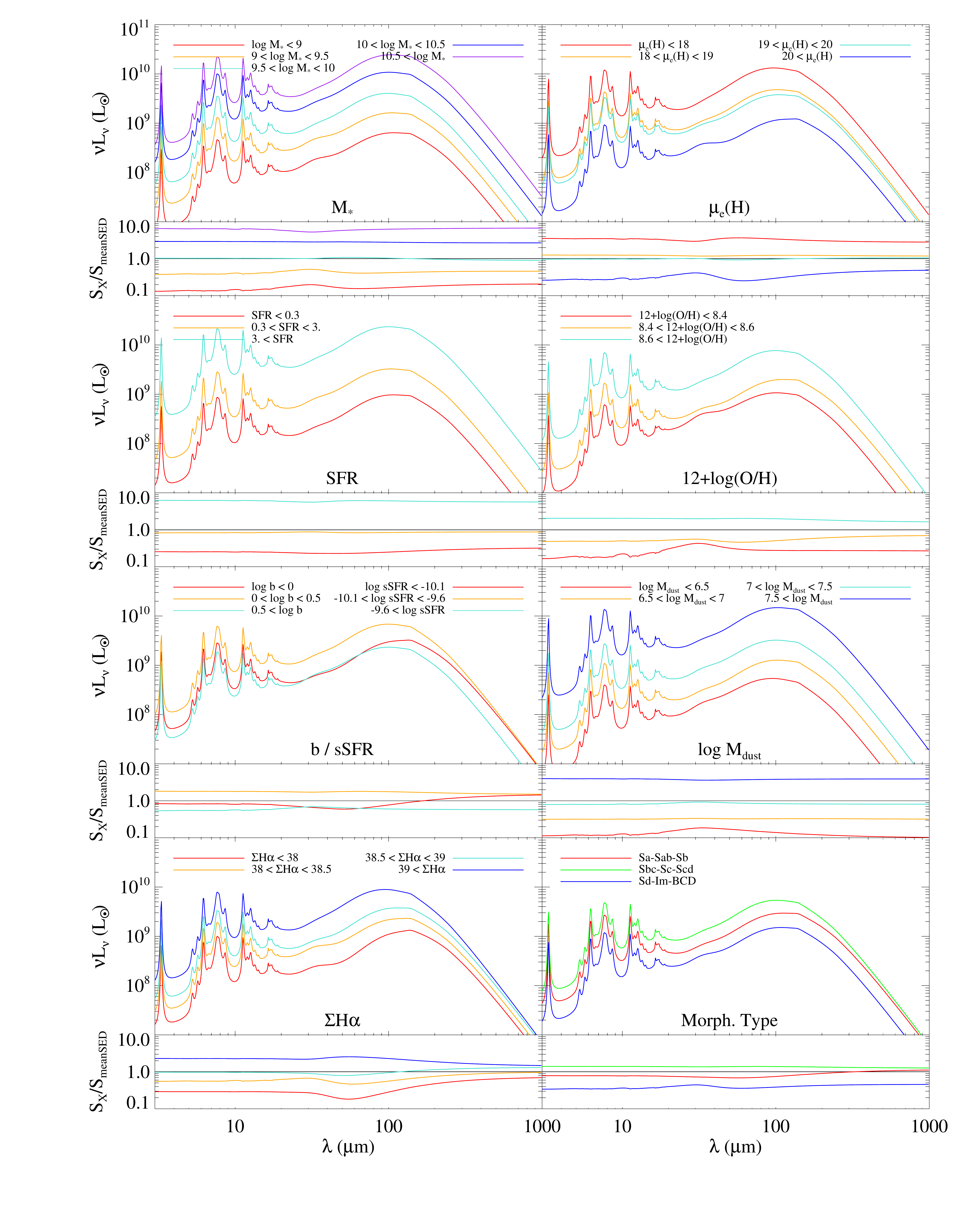}
  		\caption{ \label{templates}Library of templates derived from the fits. SEDs are binned by (from top to bottom and left to right): stellar mass, star formation rate, birthrate parameter (specific star formation rate), H$\alpha$ surface brightness, H-band effective surface brightness, metallicity, dust mass, and morphological types. For each class of templates, the lower panels show the ratio between the different binned templates and the mean HRS SED of Figure~\ref{template_mean}.}
	\end{figure*}
	
	We then binned the best fit models of the sample by birthrate parameter, star formation rate, dust mass, metallicity, stellar mass, H-band effective surface brightness, H$\alpha$ surface brightness, and morphological type (Figure~\ref{templates}).
	The bins (indicated on Figure~\ref{templates}) have been chosen in order to have a consistent number of objects in each bin.
	For each bin, the best fit models of the galaxies are averaged in order to provide the template corresponding.
	
	The new set of IR templates is constrained with submm data, and they benefit from the good photometric coverage of the HRS galaxies.
	The HRS was designed to be complete in stellar mass, and the selection criteria applied to these galaxies make our sample representative of the late-type galaxies of the local Universe.
	Therefore, these templates provide constraints on the dust emission in the nearby Universe.
	However, as these templates were built from DL07 models, we should note that they may underestimate the dust emission at $\lambda \ge$500\microns\ for low mass systems, as discussed in Section~\ref{fitresults}. 
	The mean SED template and the binned SED templates are available to the community via the HEDAM website\footnote{http://hedam.lam.fr/HRS/}.
	
	\subsection{Comparison with the literature}

	\begin{figure*}
		\includegraphics[width=10cm,angle=90]{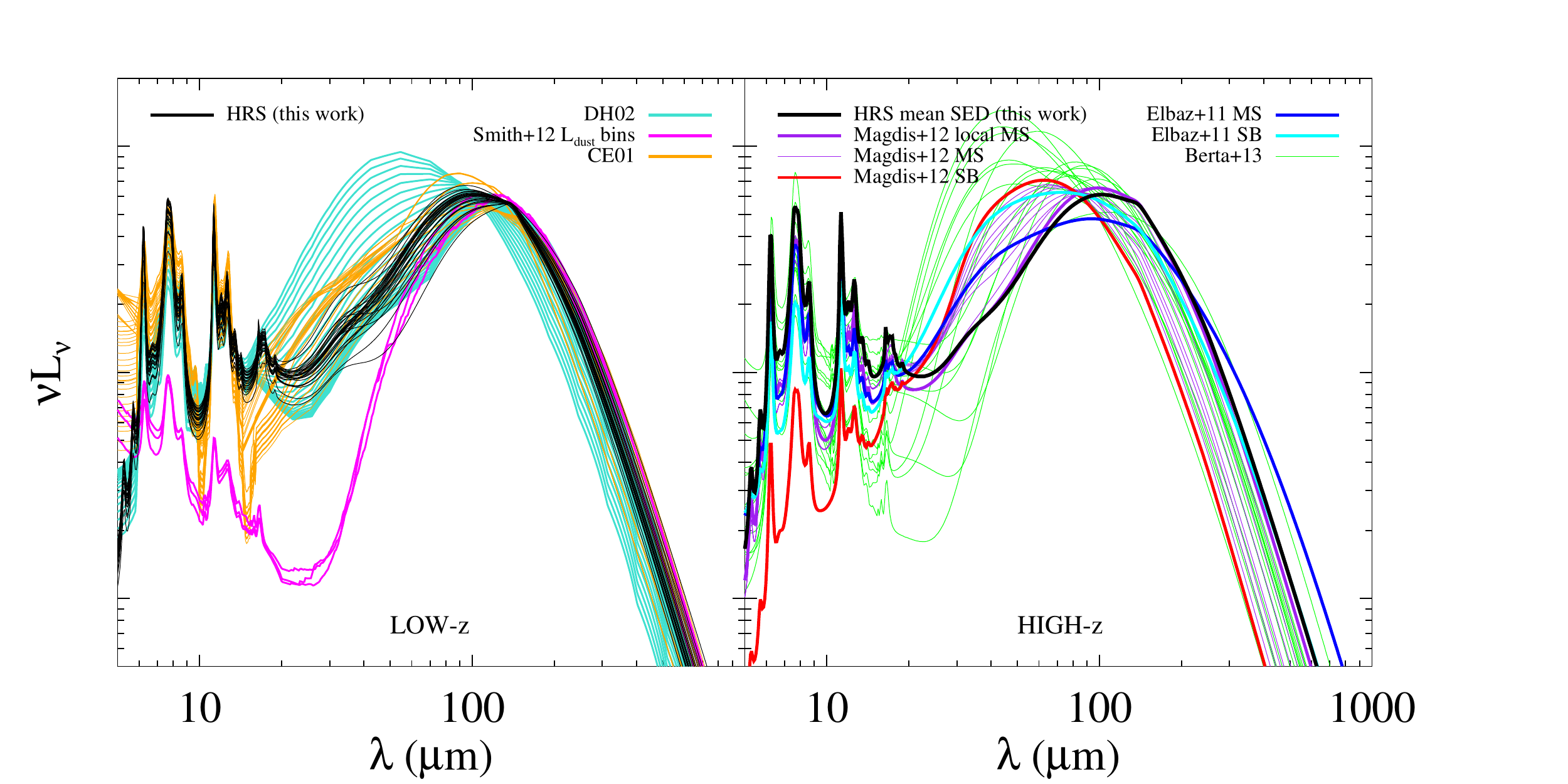}
  		\caption{ \label{comp_models} Templates from this work compared to the most popular IR SED templates normalized to $L_{8-1000}$. Left panel: Low-$z$ templates. In black all of the templates computed in this work (Figure~\ref{templates}). For better comparison, we plot only the templates of the different libraries whose properties ($L_{IR}$,$\alpha$) match those of the HRS. In turquoise: the templates of \cite{DaleHelou02} with $\alpha$ ranging from 1.5 to 4. In orange, the library of \cite{CharyElbaz01} with $\log (L_{IR}/$L$_{\odot}) < $10.9  and in magenta, the median templates of \cite{SmithDunne12} binned in dust luminosity for $\log (L_{IR}/$L$_{\odot}) < $11. Right panel: High-$z$ templates. In black: the mean SED from HRS (Figure~\ref{template_mean}) as a reference for $z$=0 galaxies. In purple: the main sequence templates and in red the starburst template from \cite{Magdis12}. In blue: the main sequence template from \cite{Elbaz11} and in cyan the starburst template from \cite{Elbaz11}. In green: selected templates from \cite{Berta13}. At low-$z$, there is a large discrepency between the templates between 20 et 100\microns. }
	\end{figure*}	
	
	We compare our templates with the most popular IR SED libraries used in the literature, i.e. CE01, DH02, \cite{Elbaz11}, \cite{Magdis12}, \cite{SmithDunne12} and \cite{Berta13}.
	These libraries are based on different physical assumptions, however this discussion is mainly focused on the shape of the SEDs  and not on the physical description of the dust.
	We thus give here a brief description of these libraries.
	
	The CE01 empirical SEDs have been computed as a function of the $L_{8-1000}$ in order to reproduce the ISO, IRAS and SCUBA data of local IR luminous galaxies.
	They used data between 0.44 and 850\microns, however there is a large gap of observations between 100 and 850\microns.
	\cite{DaleHelou02} built, using a semi-empirical method, a library of galaxy SEDs following the dust mass power law distribution as a function of the intensity of the ISRF $U$, i.e. $dM(U) \propto U^{-\alpha} dU$.
	This set of templates reproduces the IR colors of galaxies thanks to a unique parameter $\alpha$ linked to the $S_{60}/S_{100}$ flux density ratio.
	A galaxy with $\alpha \approx 1$ is active in star formation, whereas a galaxy with $\alpha \approx 2.5$ is qualified as normal.
	However, CE01 and DH02 are pre-\textit{Herschel} libraries and did not benefit from the excellent FIR/submm coverage that we have from PACS and SPIRE.
	Finally, \cite{SmithDunne12} used a sample of galaxies selected at 250\microns\ from the H-ATLAS field \citep{Eales10}.
	They used the models of  \cite{daCunha08} and performed a panchromatic SED fitting based on energy balance.
	As a result of these fits, they provided median templates binned by physical parameters for galaxies up to $z$=0.5. 
	However, we should note that they did not have any constraint between the K band and the 100\microns\ flux densities.
	
	Recent studies using \textit{Herschel} data computed libraries designed for higher redshift sources.
	Thanks to \textit{Herschel} data, \cite{Elbaz11} showed that the MS galaxies also verify a universal total to MIR luminosity ratio: $IR8=L_{IR}/L_8$, with $L_8$, the rest frame luminosity at 8 \microns, suggesting that they have the same IR SED shape. 
	In this relation, the starburst galaxies are outliers, with higher $IR8$.
	Thus \cite{Elbaz11} derived a single template for MS galaxies, with a single IR8 at all redshifts up to 2, and a single template for starburst galaxies with a significantly higher value of $IR8$. 
	\cite{Magdis12} adopted the same philosophy but introduced a variation of the shape of the SED with the redshift.
	They fit \cite{DraineLi07} models to individual galaxies and stacked ensembles at 0.5$<z<$2 using data from MIR to millimeter range.
	They derived a relation between $<U>$ and $z$ and used it to construct template SEDs of MS galaxies from $z$=0 to $z$=2.5, fixing $\gamma$=0.02 and $q_{PAH}$=3.19\% for $z<$1.5 and 2.50\% for $z>$1.5.
	They assume a flattening evolution of $<U>$ beyond $z$=2.5.
	For starburst galaxies, they used their best fit of GN\,20, a well-studied very luminous and distant submm galaxy. 
	\cite{Berta13} combined UV to submm data of galaxies from large fields (GOODS-N, GOODS-S and COSMOS) to reproduce the distribution of galaxies in 10 rest-frame color spaces, using a superposition of multi-variate Gaussian modes.
	According to this model, they classified galaxies and built the median SED of each class.
	Each median SED was then fitted using a modified version of the MAGPHYS code \citep{daCunha08} that combines stellar light, emission from dust heated by stars and a possible contribution from dust heated by an AGN. 
	
	We compare our templates with the one available in the literature in Figure~\ref{comp_models}, where they have all been normalized to the $L_{8-1000}$.
	The choice of this normalization comes from the fact that these libraries are essentially used to provide a measurement of the $L_{IR}$ of the galaxies, the templates of CE01 are even calibrated to this parameter.
	We also divide them into two categories: in the left panel, the libraries representative of the nearby Universe, and in the right panel, the libraries developed for high-$z$ studies.
	
	On the left panel, we focus first on the low-z libraries, From each library, we select the templates that match the properties of our sample.	
	For CE01, we select the templates with $\log (L_{IR}/$L$_{\odot})<$10.9, we note that there is no template for $\log (L_{IR}/$L$_{\odot})$ lower than 8.5 that could be representative of some galaxies of our sample.
	For DH02, based on the $S_{60}/S_{100}$ ratios of our sample, we select templates with $\alpha>1.5$.
	We should note that for 30\% of our galaxies having a 60\microns\ measurement, templates with $\alpha>4$, the largest value of $\alpha$ available from the DH02 library, are needed to reproduce the observed $S_{60}/S_{100}$.
	For the \cite{SmithDunne12} library, we show the templates corresponding to $\log (L_{IR}/$L$_{\odot})<$11.
	We show all of the HRS templates presented in Figure~\ref{templates}.
	CE01, DH02 and \cite{SmithDunne12} templates contain stellar emission that explain part of the differences seen at $\lambda<10$\microns\ compared to ours that are purely dust emission.
	All the libraries here are consistent within $\approx$0.2\,dex around 100\microns.
	Indeed, pre-\textit{Herschel} libraries (CE01 and DH02) benefit from IRAS 100\microns\ measurements, and post-\textit{Herschel} libraries presented here can rely on PACS data at 100\microns.
	This part of the SED is thus well constrained by the observations.
	At longer wavelength, the low-$z$ templates have consistent shapes.
	However, we remark in Figure~\ref{modelvsobs} that DL07 models have difficulties to reproduce the 500\microns\ observations, thus despite this good agreement between the different libraries, updates in the models are needed to be able to reproduce the data.
	At wavelength shorter than 100\microns, CE01 and DH02 templates represent dust temperatures that are significantly higher than those of \cite{SmithDunne12} and this work due to selection criteria.
	Indeed, the IRAS selections for the normal galaxies introduce a bias toward warmer sources.
	There is a particularly striking disagreement between the templates in MIR-FIR that can reach up to two orders of magnitude.
	The pre-\textit{Herschel} libraries have IR peaks that are shifted toward shorter wavelength compared to the post-\textit{Herschel} ones, resulting in higher temperatures for the warm dust.
	What increases the disagreement between the libraries in this 20-100\microns\ range is the very low luminosity of the \cite{SmithDunne12} median SEDs.
	This is probably due to the fact that \cite{SmithDunne12} did not have any MIR and FIR data up to 100\microns\ to constrain their templates.
	Even if we do not take into account this library, the disagreement of the low-$z$ libraries between 20 and 100\microns\ is very important.
	The emission in this range is a combination of emission from warm dust populations and stochastically heated grains.
	This warm dust which peaks around 100\microns\, and even at shorter wavelengths, is theoretically an excellent tracer of star formation.
	It is unclear if the larger scatter we observe in FIR is only due to uncertainties on the way templates are built (there are few observational constraints in this range), or whether this translates variations in the physical properties in galaxies.
	Indeed, based of the dust emission models of \cite{DraineLi07}, the relative contribution of photo-dissociation regions and diffuse stellar component to the heating of the dust impacts the SED in this particular range.
	
	On the right panel, we show our mean SED as a representative SED for $z$=0 galaxies along with templates based on the study of high-$z$ galaxies done with \textit{Herschel} observations.
	As for the low-$z$ libraries, these templates agree relatively well around 100\microns, except for the SED of the MS galaxies from \cite{Elbaz11}. 
	At longer wavelengths, the high-$z$ library of \cite{Magdis12} and of this work have also the same slope.
	\cite{Elbaz11} computed their SEDs template by meaning all of the galaxies SEDs of their sample, normalized to $10^{11}/$L$_{IR}$, resulting in a lower $\beta$ for both their MS and starburst templates.
	The $z$=0 main sequence template of \cite{Magdis12} (thick purple solid line) is in good agreement with the mean SED of HRS galaxies.
	With the increase in redshift, \cite{Magdis12} templates shift toward shorter wavelength, and thus higher dust temperature.
	We select some templates out of the 32 available from the library of \cite{Berta13} by removing templates with an AGN components and those typical of high-$z$ sources, such as Lyman break galaxy template.
	These templates show a large variety of SED shape, temperature of the dust and PAH bands intensity.
	The submm part of the \cite{Berta13} templates is very similar to \cite{Magdis12} and the mean SED of this work, however the MIR-FIR domain show a large range of luminosities.

\section{Conclusions}
Thanks to the wealth of photometric ancillary data available for the HRS, we perform the NIR photometry of the HRS galaxies (see Appendix~\ref{photom}) and compute their IR SED from 8 to 500\microns\.
We provide an updated method to remove the stellar contribution in NIR and MIR using different SFH according to the morphological type of the galaxies.
The \cite{DraineLi07} models are fit to the HRS galaxies.
Even though this work focuses on the dust properties of a gas-rich galaxy subsample, we provide the results of the SED fitting of the other HRS galaxies (early-type and H{\sc I}-deficient galaxies) in Appendix~\ref{defparam}.
We note that a strong constraint in the 20-60\microns\ range, at least a reliable measurement at 24\microns, is mandatory to have a reliable estimation of the relative contribution of PDR to the total IR SED.
The comparison between observed and modeled flux densities shows that \cite{DraineLi07} models underestimate the 500\microns\ observations for low mass systems.
We also note an underestimation of the 160\microns\ measurements, especially for high-mass systems.

The median diffuse ISRF intensity of our late-type sample is 1.96$\pm$0.47 times the ISRF intensity of the Milky Way corresponding to a median dust temperature of 22\,K, with a median contribution of the PDRs of 0.75\%$\pm$0.5\%, and a contribution of the PAH to the total dust mass of 4.16\%$\pm$0.42.
The median $\log (L_{IR}/$L$_{\odot})$ is 9.47$\pm$0.03 and the median dust mass of the sample is $\log (M_{dust}/$M$_{\odot})$7.02$\pm$0.08.

We compared these parameters derived from the fitting procedure to integrated properties of the galaxies.
From this comparison, in agreement with \cite{Boselli12}, we confirm the good correlation between the fraction of PAH and the metallicity implying a weakening of the emission of the PAH in galaxies with lower metallicities. 
We thus confirm results obtained from MIR colors studies \citep{Boselli04,Engelbracht05} and spectroscopy \citep{Smith07}. 
From this relation follows a good correlation between the fraction of PAH and the stellar mass, as most massive systems have higher metallicities.
Moderate to good correlations are found between the minimum intensity of the ISRF and the H-band and H$\alpha$ surface brightness implying that, based on integrated galaxy analysis, the diffuse dust component seems to be heated by both the young stars in star forming regions and the diffuse evolved populations. 
This confirms the results of \cite{Boselli12} also based on integrated study of galaxies but is incompatible with analysis based on resolved studies.

We placed the HRS galaxies in a SFR-$M_*$ diagram, and compare them with fits of the MS galaxies at different redshifts.
There is a good agreement with the MS relation at $z=0$ of \cite{Peng10}, even if HRS low mass systems tend to have higher SFR.
The best linear fit to the data is $\log SFR = 0.65 \times \log M_* - 6.29$.
Thanks to the good photometric coverage of our sample, we are able to provide IR templates of nearby galaxies binned by the parameters that constrain the shape of the IR SED: birthrate parameter (or equivalently the $sSFR$), dust mass, metallicity, stellar mass, H-band effective surface brightness, H$\alpha$ surface brightness, morphological type.
We also computed the mean SED of the subsample in order to provide a reference for any cosmological studies and simulations.
The mean SED and the library can be found on the HEDAM website\footnote{http://hedam.oamp.fr/HRS/}.
We compared this set of templates to the most popular IR pre-\textit{Herschel} and post-\textit{Herschel} libraries.
At low-$z$, pre-\textit{Herschel} libraries have a very warm dust component compared to post-\textit{Herschel} ones, due to selection effects of IRAS galaxy samples.
Compared to high-$z$ libraries, our mean SED is in good agreement with the $z=0$ MS template of \cite{Magdis12}.
The IR peak of the MS template of \cite{Elbaz11} is wider than the IR peak obtained in this work, and corresponds to a lower luminosity at 160\microns.
In the Appendix, we provide new photometric data in the 8, 12, and 22\microns\ bands taken by \textit{Spitzer} and WISE used in the work.
We also present in the Appendix, new coefficients to remove the stellar contribution from MIR observations.

\begin{acknowledgements}
We thank the anonymous referee for his/her comments which greatly helped improving this paper.
LC thanks Daniel Dale for sharing KINGFISH results and discussions and Vassilis Charmandaris for useful comments and discussions.
IDL is a postdoctoral researcher of the FWO-Vlaanderen (Belgium).
SPIRE has been developed by a consortium of institutes led
by Cardiff Univ. (UK) and including Univ. Lethbridge (Canada); NAOC (China);
CEA, LAM (France); IFSI, Univ. Padua (Italy); IAC (Spain); Stockholm
Observatory (Sweden); Imperial College London, RAL, UCL-MSSL, UKATC,
Univ. Sussex (UK); Caltech, JPL, NHS C, Univ. Colorado (USA). This development
has been supported by national funding agencies: CSA (Canada); NAOC
(China); CEA, CNES, CNRS (France); ASI (Italy); MCINN (Spain); SNSB
(Sweden); STFC, UKSA (UK); and NASA (USA). This research has made
use of the NASA/IPAC ExtraGalactic Database (NED) which is operated by
the Jet Propulsion Laboratory, California Institute of Technology, under contract
with the National Aeronautics and Space Administration. This research has made use of the NASA/IPAC ExtraGalactic
Database (NED) which is operated by the Jet Propulsion Laboratory, California
Institute of Technology, under contract with the National Aeronautics and Space
Administration and of the GOLDMine database (http://goldmine.mib.infn.it/).
This publication makes use of data products from the Wide-field Infrared Survey Explorer, 
which is a joint project of the University of California, Los Angeles, and the Jet Propulsion 
Laboratory/California Institute of Technology, funded by the National Aeronautics and Space Administration.

\end{acknowledgements}


\bibliographystyle{aa}
\bibliography{hrs_sed}

\begin{appendix}
\section{\label{photom} MIR photometry of the HRS}

	\begin{table}
	\centering
	\caption{Statistics of the MIR photometry from IRAC and WISE.}
	\begin{tabular}{c c c c}
  	\hline\hline
	  &\textit{Spitzer}/IRAC  & WISE &WISE\\
  	  &8\microns\  & 12\microns\ &22\microns\\
  	\hline 
  	Number of galaxies 	& 129	       & 323		& 323  \\ 
  	Detections 			& 91$\%$	& 96$\%$ 	& 86$\%$ \\ 
	Mean error 			& 15$\%$	& 6$\%$		& 13$\%$  \\ 
	\hline
	\label{MIRstat}
	\end{tabular}
	\end{table}
	
For the purpose of this study, we perform MIR photometry at 8, 12 and 22\microns\ from \textit{Spitzer}/IRAC and WISE data.
We thus present here the flux densities at these wavelengths for all the HRS galaxies.

	\subsection{\textit{Spitzer}/IRAC 8\microns}
	
		One of the output results from \cite{DraineLi07} models is the PAH fraction of the total dust mass of a galaxy.
		Because one of the largest PAH feature emission is expected at 7.7\microns, we need the IRAC \citep{Fazio04} 8\microns\ images to constrain this part of the SED.
	
		We retrieved from the \textit{Spitzer} Data Archive\footnote{\url{http://irsa.ipac.caltech.edu/data/SPITZER/docs/spitzerdataarchives/}}, 8\microns\ images available for 129 HRS galaxies.
		The FWHM of the PSF of the IRAC fourth channel is 1.9\arcsec\, and the maps have a pixel size of 0.6\arcsec.
		As a first step, we convert the images from MJy/sr to $\mu$Jy/pixel, and then remove all of the problematic pixels (NaN) by replacing them by the median value of the surrounding pixels.
		We remove the stars and background sources visible at 8\microns\ using the \texttt{IRAF/imedit} task from all of the images.
		We then extract the flux densities using apertures adapted to each galaxies in order to take into account all of the IR emission, and minimize the contamination from background features.
		As the images have small fields of view, we do not estimate residual background emission from circular annuli (as we do for WISE and \textit{Herschel}/SPIRE photometry) but from the mean value of multiple 10$\times$10 pixels boxes around the galaxies.
		IRAC flux densities need to be corrected for aperture effects to take into account the diffuse scattering of incoming photons through the IRAC array\footnote{\url{http://ssc.spitzer.caltech.edu/irac/calib/extcal/}}.
		We thus apply aperture corrections using the parameters provided in Table 4 of \cite{Dale07} on our measurements.
		
		For the error calculation, we proceed as in \cite{Boselli03b} using the same boxes as for the background residual estimation.
		We thus take into account two terms, one being the pixel-to-pixel error (the mean value of the standard deviation in all of the boxes) and the other one the sky error due to large scale structures (the standard deviation of the mean values in all of the boxes).
		To this stochastic error, we add quadratically a calibration error of 10\% as indicated in \cite{Dale07}. 
		Galaxies with a signal-to-noise lower than 3 are considered as undetected, and an upper limit of 3$\sigma$ is given.
		
		In Table~\ref{MIRflux}, we give the IRAC 8\microns\ flux densities of the HRS galaxies.
		A flag is associated to the measurements with the following code: 0 for undetected galaxies, 1 for detected galaxies and 2 for galaxies for which the measurement suffers from source blending.
	
		\begin{figure}
 			\includegraphics[width=6cm,angle=90]{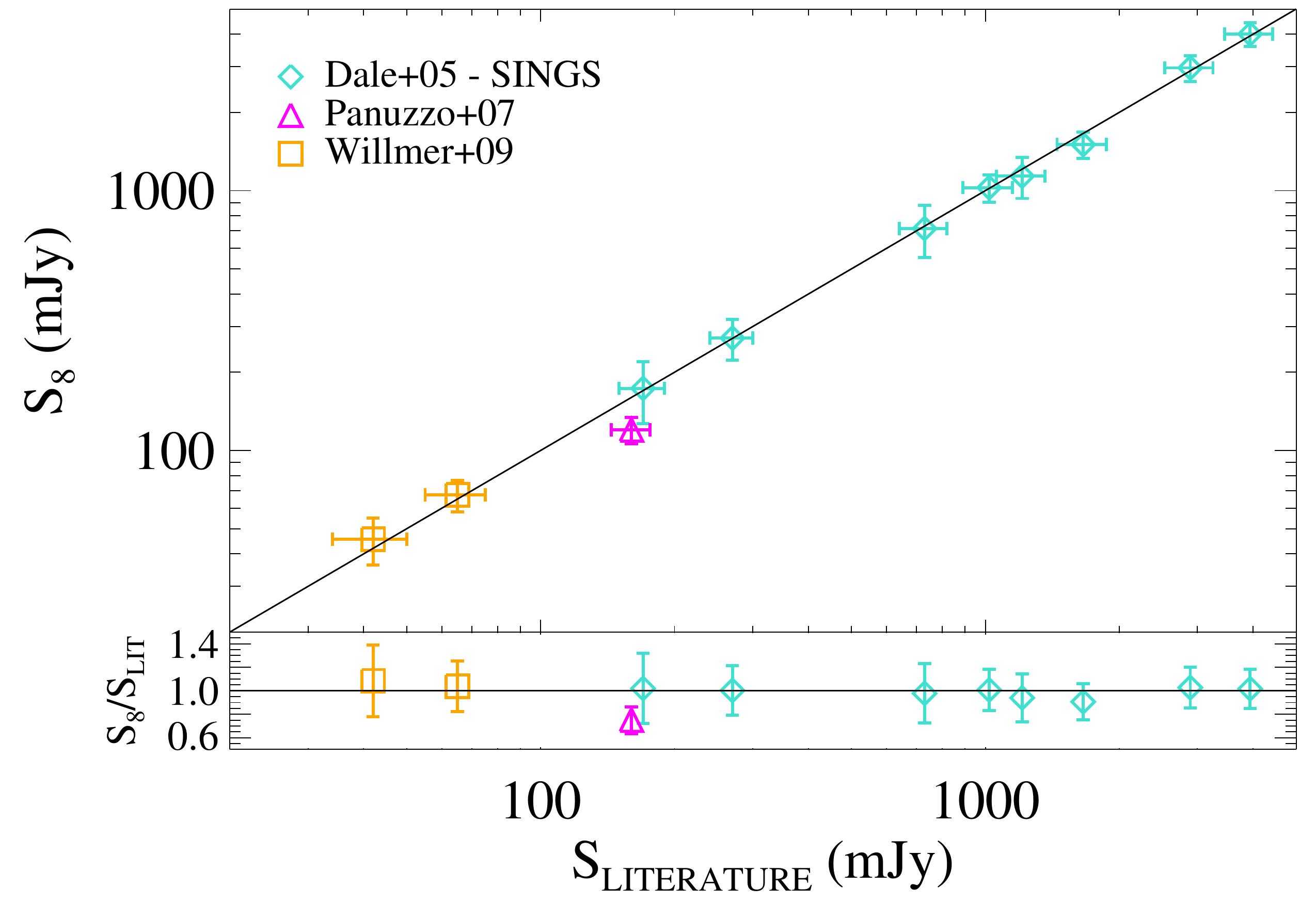}
  			\caption{ \label{compirac} Comparison between our \textit{Spitzer}/IRAC 8\microns\ flux density measurements and  \textit{Spitzer}/IRAC data from the literature. The one-to-one relationship is the solid black line. The ratios between the HRS IRAC 8\microns\ and the literature 8\microns\ flux densities are shown in the lower panel. }
 		\end{figure}
	
		In order to check the validity of our IRAC photometry, we search for IRAC 8\microns\ flux densities already available for HRS galaxies on NED and make the comparison (Figure~\ref{compirac}).
		We find 8 galaxies in common with the SINGS sample \citep[the \textit{Spitzer} Infrared Nearby Galaxies Survey,][]{Kennicutt03,Dale07}, 1 galaxy in common with \cite{Panuzzo07} and 2 galaxies to be compared with \cite{Willmer09}.
		The only point that deviates from the one-to-one relationship is the data point of \cite{Panuzzo07}.
		It corresponds to the flux density of NGC\,4435 which is very close to NGC\,4438.
		This proximity can be problematic as outlined by \cite{Panuzzo07} and the difference in the measure may come from how the emission from NGC\,4438 is treated.
		This comparison shows that our measurements are consistent with the 8\microns\ data taken from the literature with a mean ratio of 0.93$\pm$0.04.

	\subsection{WISE 12 and 22\microns}
		
		The NASA's Wide-field Infrared Survey Explorer \citep[WISE,][]{Wright10} performed an all sky survey in four NIR and MIR bands.
		For the purpose of this work, we use WISE data at 12 and 22\microns, observed with a resolution of 6.5\arcsec, and 12.0\arcsec, respectively.
		WISE scanned the sky with 8.8 seconds exposures at 12 and 22\microns\  (the W3 and W4 bands), each with a 47$\arcmin$ field of view, providing at least eight exposures per position on the ecliptic and increasing depth towards the ecliptic poles. 
		The individual frames were combined into coadded images with a pixel size of 1.375\arcsec. 
		WISE achieved $5\sigma$ point source sensitivities better than 1 and 6 mJy in unconfused regions on the ecliptic in the W3 and W4 bands. 
		Sensitivity improves toward the ecliptic poles due to denser coverage and lower zodiacal background.
		We retrieve the images of all of the HRS galaxies at 12 and 22\microns\ from the WISE Science Archive\footnote{Science Archive:\url{http://irsa.ipac.caltech.edu/}}.
		We perform aperture photometry using the DS9/Funtools program ''Funcnts''.
		For each galaxy, the aperture, where the flux of the galaxy is estimated, and the background annulus, where the emission from the background is estimated, are defined ``manually''.
		They are chosen in order to encompass all of the emission from the galaxy and avoid any contamination from foreground/background sources.
		
		To convert the counts extracted into Jy, we use the factors provided by the Explanatory Supplement to the WISE Preliminary Data Release Products\footnote{\url{http://wise2.ipac.caltech.edu/docs/release/prelim/expsup/wise_prelrel_toc.html}}, Section II.3.f, $2.9045\times10^{-6}$\,Jy/DN and $5.2269\times10^{-5}$\,Jy/DN at 12 and 22\microns, respectively.
		For WISE photometry of extended sources, \cite{Jarrett13} recommended 3 corrections to be applied to all measurements.
		The first one is an aperture correction that accounts for the WISE absolute photometric calibration method using PSF profile fitting.
		This correction is 0.03\,mag and -0.03\,mag at 12 and 22\microns, respectively.
		The second correction is a color correction that accounts for the spectral signature of the source convolved with WISE relative system response (RSR).
		Because our SED fitting method integrates the models into the filters of the bands before comparing them with the data, we do not need to apply this correction.
		The third correction comes from a discrepancy between the WISE photometric standard ``blue'' stars and ``red'' galaxies related to an error in the W4 RSR, as described in \cite{Wright10} and \cite{Jarrett11}.
		Following \cite{Jarrett13}, we apply a correction factor of 0.92 to the 22\microns\ flux densities of all of the spirals and disk galaxies.
		This correction is thus applied to HRS galaxies with morphological type of Sa and later.
		
		We determine the errors on the measurements following the method described for the IRAC data.
		However, because of a correlated noise with a typical length scale larger than 10 pixels, we use 50$\times$50 boxes rather than 10$\times$10 pixels.
		Galaxies with a signal-to-noise lower than 3 are considered as undetected, and an upper limit of 3$\sigma$ is given.
		
		In Table~\ref{MIRflux}, we give the WISE 12 and 22\microns\ flux densities of the HRS galaxies.
		As for IRAC measurements, a flag is provided with the following code: 0 for undetected galaxies, 1 for detected galaxies and 2 for galaxies for which the measurement suffer from sources blending.
		
		\begin{figure*}
 			\includegraphics[width=9cm]{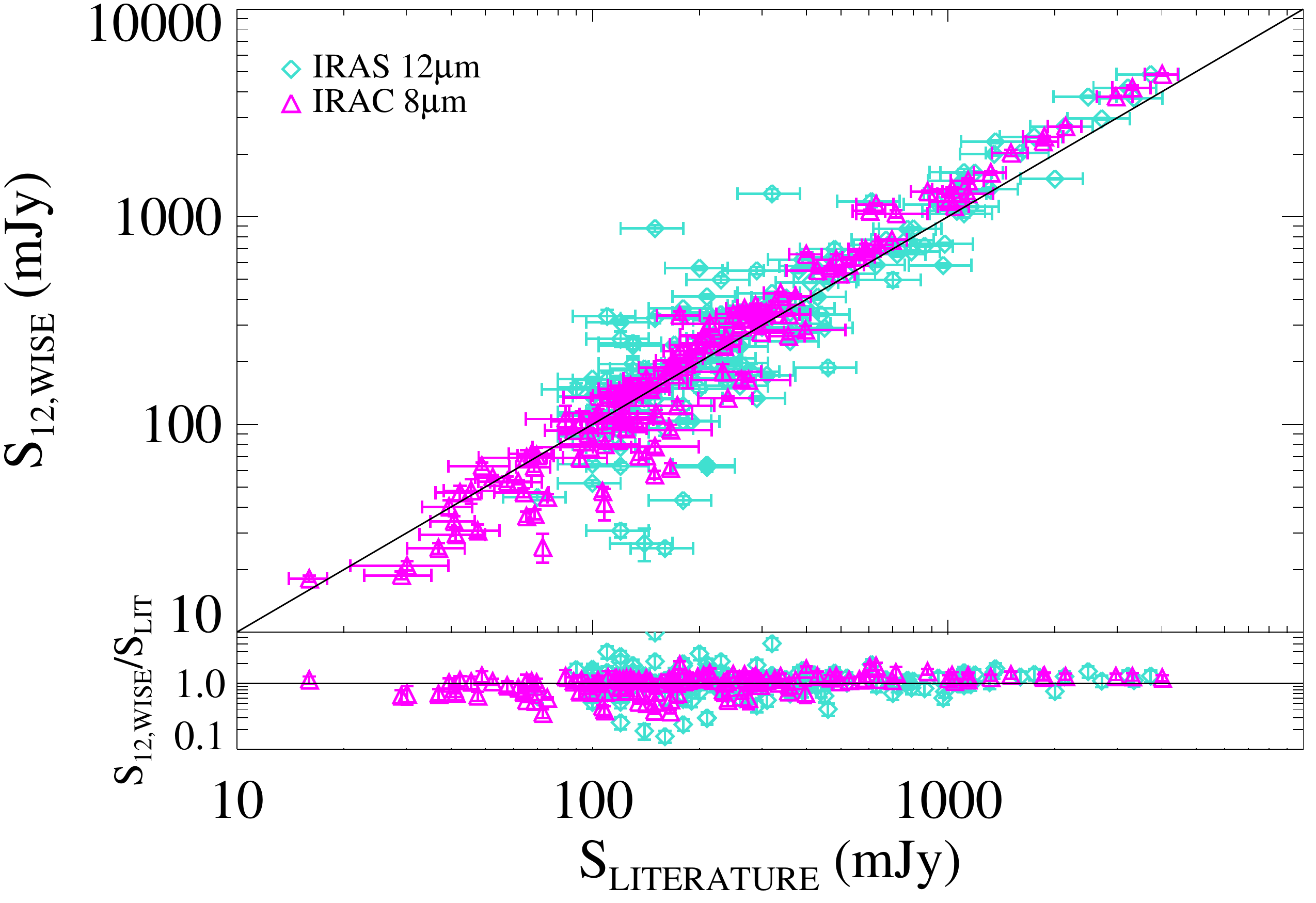}\hfill
 			\includegraphics[width=9cm]{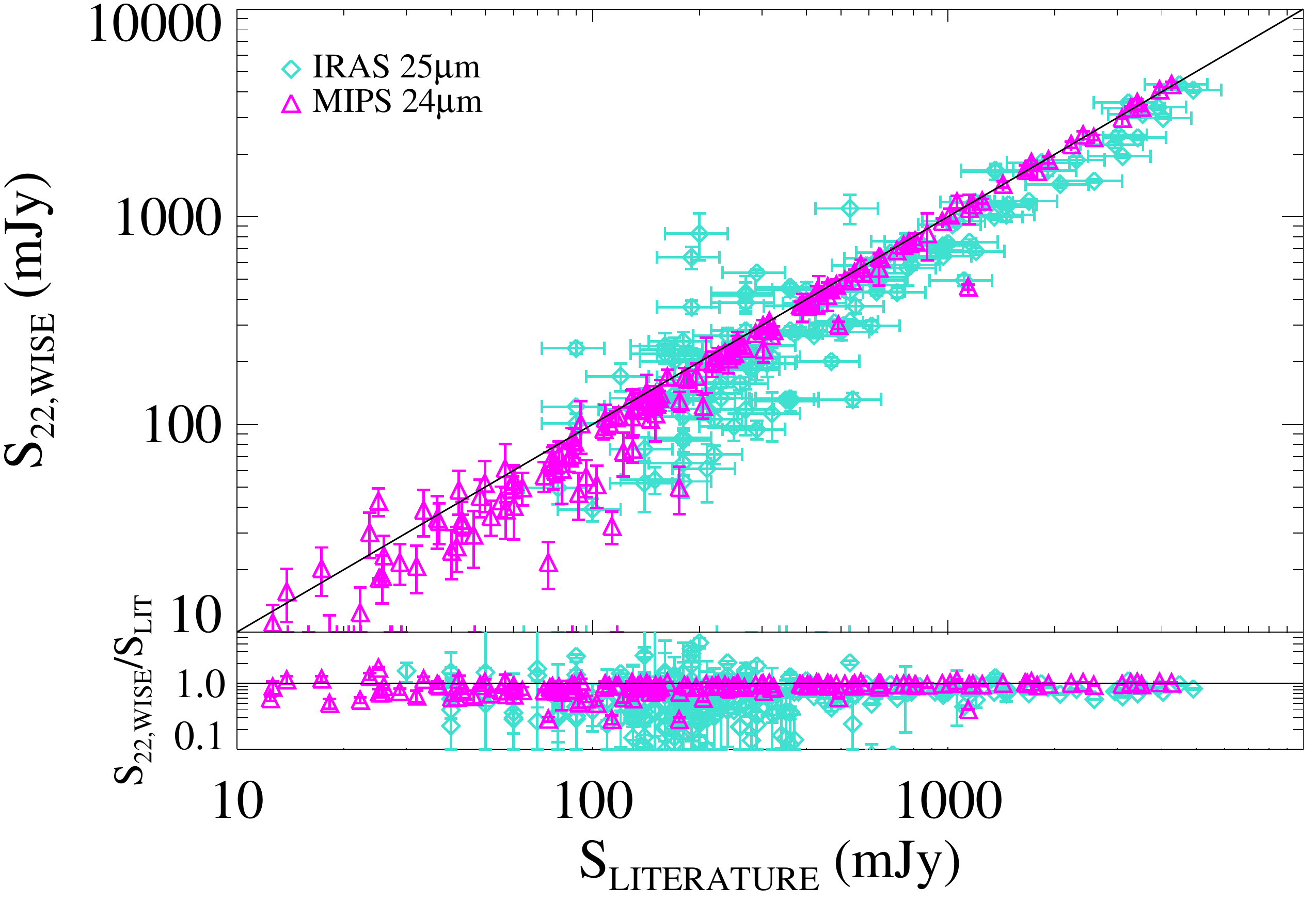}
  			\caption{ \label{compwise} Comparison between WISE flux density measurements and NIR/MIR ancillary data available from the literature at 12 and 22\microns\ (left and right panel, respectively). The one-to-one relationship is the solid black line. The WISE to literature flux density ratios are shown in the lower panels.}
 		\end{figure*}

		To check the validity of our measurements, we compared our results to the NIR and MIR ancillary data available for the HRS galaxies in Figure~\ref{compwise}.
		We use the \textit{Spitzer}/IRAC 8\microns\ from this work and the \textit{Spitzer}/MIPS 24 of \cite{Bendo12a}. 
		Even if the 8\microns\ filter from IRAC and the 12\microns\ filter from WISE are not overlapping enough to make a reliable comparison, and knowing that the emission process at the two wavelengths differ, we plot the comparison in order to identify possible outliers and thus possible issues with the photometry.
		IRAS data at 12 and 25\microns\ are also available for some of the HRS galaxies  \citep{Sanders03,Moshir90,ThuanSauvage92,Soifer89,Young96}. 
		Some shifts in the relation can be due to the different wavelength of the data and the different response curve of the filters.
		There is a good correlation between IRAS and WISE data at 12\microns\ but WISE flux densities tend to be higher than IRAS ones. 
		The mean $S^{12}_{IRAS}/S^{12}_{WISE}$ is 1.11 with a standard deviation of 0.79.
		We note that the relation becomes very dispersed for flux densities below $\approx$80\,mJy.
		The relation between the IRAC 8\microns\ and the WISE 12\microns\ measured in this work is good, with a standard deviation of 0.43 for the IRAC to WISE flux density ratio, and a mean value of the ratio of 1.08.
		There might be an effect due to the difference of wavelength, however it is difficult to quantify it as the 8\microns\ lies completely in the PAH emission domain.
		
		There is good agreement between MIPS 24\microns\ and WISE 22\microns\ measurements with a mean MIPS to WISE flux density ratio of 1.22$\pm$0.44. 
		Even if very dispersed, the relation between IRAS 25\microns\ and WISE 22\microns\ flux densities is good.
		However, as noticed at 12\microns, the relation becomes very dispersed for $S_{25}^{IRAS}<~$80\,mJy. 
\tiny
\onecolumn{
\begin{landscape}
\begin{longtable}{c c c c c c c c c c c c c c c c c }
\caption{\label{MIRflux} MIR photometry of the HRS galaxies.} \\
\hline
\hline
HRS  & CGCG   &  VCC   & UGC  &  NGC  &  IC   &  Flag 8\microns\ &   $S_{8}$ (mJy) & Flag 12\microns\ & $S_{12}$ (mJy) & Flag 22\microns\ & $S_{22}$ (mJy) & RA$_{ap}$ & Dec$_{ap}$ & a (") & b (") & PA (deg)  \\
\hline
 \endfirsthead
\caption{continued.}\\  
\hline
\hline
HRS  & CGCG   &  VCC   & UGC  &  NGC  &  IC   &  Flag 8\microns\ &   $S_{8}$ (mJy) & Flag 12\microns\ & $S_{12}$ (mJy) & Flag 22\microns\ & $S_{22}$ (mJy) & RA$_{ap}$ & Dec$_{ap}$ & a (") & b (") & PA (deg)  \\
\hline
\endhead
\hline
\endfoot
       1       &    123035       &         -       &         -       &         -       &         -       &         0           & - &         1       &           29.47   $\pm$            2.46       &         1        &          21.24   $\pm$            3.56       &          154.41       &           22.81       &            47.9       &            40.1       &    260       \\
       2       &    124004       &         -       &      5588       &         -       &         -      &         0           & - &         1       &           89.90   $\pm$            2.08       &         1        &          78.96   $\pm$            7.19       &          155.24       &           25.36       &            40.1       &            37.4       &    130       \\
       3       &     94026       &         -       &      5617       &      3226       &         -       &         0           & - &         1       &           57.34   $\pm$            2.01       &         1        &          42.69   $\pm$            6.57       &          155.86       &           19.90       &            39.4       &            39.0       &    105       \\
       4       &     94028       &         -       &      5620       &      3227       &         -       &         1       &          628.01   $\pm$           75.06       &         1       &         1145.72   $\pm$           16.37       &         1        &        1823.09   $\pm$           35.59       &          155.88       &           19.86       &           100.5       &            82.9       &    245       \\
       5       &     94052       &         -       &         -       &         -       &       610       &         0           & - &         1       &          114.02   $\pm$            3.28       &         1        &         108.32   $\pm$           10.98       &          156.62       &           20.23       &            62.8       &            38.2       &    118       \\
\hline
\end{longtable}
\end{landscape}
}
\twocolumn
\normalsize

\section{\label{stell} Removing the stellar contribution}

	In global galaxies, the emission in the 2-10\microns\ range is due to both the old stellar population and the dust (very small grains + PAHs).
	The contribution of these two components varies differently with the type of the galaxies, the former being dominant in ETG, the latter in star forming systems.
	Given that the \cite{DraineLi07} models deal only with the dust component, we first have to remove the stellar contribution.
	This has been historically done by considering the Rayleigh-Jeans tail of the stellar emission peaked at $\approx$1-2\microns\ determined using a blackbody with $T\approx$3000K, using the typical SED of ETG not showing any evidence of the presence of dust, or using stellar SEDs derived, for instance, from the Starburst99 models \citep[e.g.,][]{Boselli98,Helou04,Draine07}.
	In this work, we use the CIGALE code \citep{Noll09} which allows us to compute the stellar SED of a galaxy using stellar population models from \citep{Maraston05} convolved with a given star formation history (SFH). 
	We refer the reader to \cite{Noll09} for a complete description of the code.

	\begin{figure}
 		\includegraphics[width=\columnwidth]{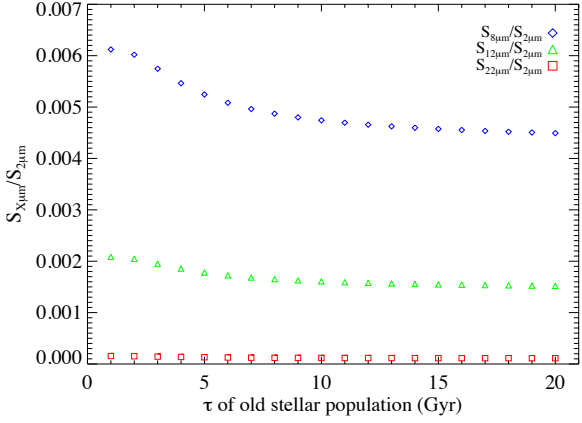}
  		\caption{ \label{tau} Evolution of $S_{8 \mu m}/S_{2 \mu m}$, $S_{12 \mu m}/S_{2 \mu m}$ and $S_{22 \mu m}/S_{2 \mu m}$ flux density ratios with $\tau$, the e-folding rate of the exponentially decreasing SFR.}
 	\end{figure}

	\begin{figure}
 		\includegraphics[width=\columnwidth]{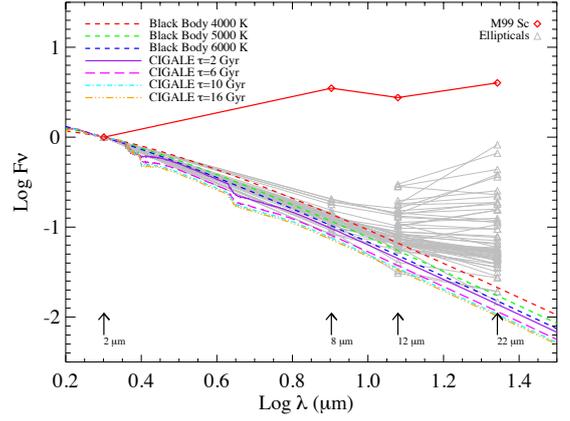}
  		\caption{ \label{constell} SED decomposition. Red diamonds are flux densities of the late type galaxy M~99 normalized to 2\microns. Grey triangles are flux densities of all HRS early-type galaxies normalized to 2\microns. Red, green and blue dashed lines represents black body laws with different temperatures. Purple, magenta, cyan and orange lines represents models from \cite{Maraston05} convolved by CIGALE with an exponentially decreasing SFH for different e-foldings $\tau$.}
 	\end{figure}	
	
	Depending on their morphological types, the stellar populations of galaxies have very different characteristics.
	The elliptical and lenticular galaxies (ETG) are dominated by an old stellar population emitting in NIR whereas stellar populations of late-type galaxies are younger, emitting in UV.
	In order to properly remove the stellar contribution from the MIR emission, we need to take these differences into account.
	
	Here, we consider a SFH with a decreasing exponential shape and an e-folding time $\tau$.
	If the e-folding time of the SFH is very small (1-2\,Gyr) then all of the stars are created in a very short time and evolve to become, at the age of the Universe, a very old population.
	On the contrary, a large e-folding ($\approx$10-20\,Gyr) corresponds to a quasi-constant SFR and stars are still being created at the age of the Universe.
	
	In order to understand the impact of the choice of $\tau$ on MIR colors, we show on Figure~\ref{tau} the variation of MIR flux densities from the models normalized to 2\microns\ with $\tau$.
	Strong variations are seen for the 8 to 2\microns\ flux density ratios and the 12 to 2\microns\ flux density ratios (36\% and 40\%, respectively), whereas a very weak variation is seen at 22\microns.
	These variations stress the fact that estimates of the stellar emission in MIR strongly depends on the SFH of the galaxy and thus on its morphological type.
	Small values of $\tau$ correspond to a high $S_{8 \mu m}/S_{2 \mu m}$ ratio, which is consistent with an important old stellar population.
	Thus, choosing a $\tau$ of 2\,Gyr would lead to a SFH compatible with ETGs.
	With $\tau$=10\,Gyr, there is less variation in the 8 to 2\microns\ flux density ratio.
	This smaller ratio implies a smaller flux density at 8\microns\ which is what we expect for late-type galaxies.
	
	To estimate the counterpart of the MIR emission due to stars in a late-type galaxy, \cite{Helou04} used the stellar population models of Starburst 99 \citep{Leitherer99}.
	Assuming that the 3.6\microns\ emission was purely stellar, they obtained stellar contribution factors of 0.596, 0.399, 0.232 and 0.032 at 4.5, 5.8,8.0 and 24.0\microns, respectively.
	\cite{Draine07} proposed similar values obtained from a blackbody emission with a temperature of 5000\,K (0.260, 0.0326 at 8.0 and 24.0\microns, respectively).
	We use blackbodies of different temperatures and stellar population models computed from CIGALE and present the resulting stellar emission normalized at 2\microns\ in Figure~\ref{constell}.
	For comparison, we show the normalized observed flux densities of all the early-type galaxies of the HRS as well as those of M\,99, a typical late-type galaxy of our sample.
	Differences in the stellar contributions are seen depending on the model (blackbody or models from CIGALE) and on the assumption on the temperature or the e-folding time.
	Both methods, black body or more complex models, pull uncertainties due to the assumption on the parameters (temperature or e-folding time).
	However, using the models from CIGALE allows us to take into account the observed differences between the stellar populations of the ETGs and LTGs. 
	We thus decide to use the stellar population models computed by CIGALE, with a population aged of 13\,Gyr, and consider these differences by using different e-folding times: $\tau=2\,Gyr$ for the ETGs and  $\tau=10\,Gyr$ for the LTGs.
	In Table~\ref{stellcon}, we calculate the coefficients and the errors associated corresponding to the stellar contribution for several NIR and MIR bands (from J to IRAS 60\microns), normalized to different bands (from J to IRAC\,1).
	Our coefficients are in good agreement with those of \cite{Helou04} and \cite{Draine07}, i.e. 0.589, 0.396, 0.244 and 0.044 at 4.5, 5.8, 8.0 and 24.0\microns, respectively for the late-type galaxies.

\begin{table*}[width=\columnwidth]
	\centering
	\caption{\label{stellcon} Stellar contribution in near- and mid-infrared bands for early-type galaxies (ETG) and late-type galaxies (LTG).}
	\begin{tabular}{c c c c c c c} 
	\hline
	\hline
			&	Band & $\lambda$ (\microns\ ) &  \multicolumn{4}{c}{Normalisation} \\
			&		& 						 &  to 1.2\microns\ (J) & to 1.6\microns\ (H) & to 2.2\microns\ (K)	& to 3.6\microns\ (IRAC 1) \\
	\hline
	
	ETG	&	J	& 1.2 		& 1. 						& -									& -			& -			\\ \\
			&	H	& 1.6		& 1.177$^{+0.003}_{-0.004}$	& 1.									& 	-		& -			\\ \\
			&	K	& 2.2		& 0.858$^{+0.002}_{-0.005}$	& 0.728$^{+0.0}_{-0.007}$			& 1.			& -			\\ \\
			&	IRAC1	& 3.6	& 0.414$^{+0.007}_{-0.036}$	& 0.352$^{+0.005}_{-0.029}$			& 0.483$^{+0.007}_{-0.044}$	& 1.			\\ \\
			&	IRAC2	& 4.5	& 0.247$^{+0.005}_{-0.023}$	& 0.210$^{+0.004}_{-0.018}$			& 0.288$^{+0.005}_{-0.028}$	& 0.596$^{+0.002}_{-0.004}$	\\ \\
			&	IRAC3	& 5.8	& 0.167$^{+0.004}_{-0.016}$	& 0.142$^{+0.003}_{-0.013}$			& 0.194$^{+0.004}_{-0.019}$	& 0.402$^{+0.001}_{-0.003}$	\\ \\
			&	IRAC4	& 8		& 0.103$^{+0.002}_{-0.010}$	& 0.088$^{+0.002}_{-0.008}$			& 0.120$^{+0.002}_{-0.012}$	& 0.249$^{+0.001}_{-0.002}$	\\ \\
			&	WISE3	& 12	& 0.083$^{+0.002}_{-0.008}$	& 0.071$^{+0.001}_{-0.006}$			&  0.097$^{+0.002}_{-0.010}$	& 0.201$^{+0.001}_{-0.002}$	\\ \\
			&	WISE4	& 22	& 0.018$^{+0.0004}_{-0.002}$	& 0.015$^{+0.0003}_{-0.001}$		& 0.021$^{+0.0004}_{-0.002}$	& 0.044$^{+0.0002}_{-0.001}$	\\ \\
			&	MIPS1	& 24	& 0.019$^{+0.0004}_{-0.002}$	& 0.016$^{+0.0003}_{-0.002}$		& 0.022$^{+0.0005}_{-0.002}$	& 0.046$^{+0.0002}_{-0.001}$	\\ \\
			&	IRAS	& 60	& 0.004$^{+0.0001}_{-0.0004}$	& 0.003$^{+0.0001}_{-0.0003}$	& 0.004$^{+0.0001}_{-0.0005}$	& 0.009$^{+0.00006}_{-0.0001}$	\\ \\
	\hline
	
	LTG	&	J	& 1.2 		& 1. 						& -									& -			& -			\\ \\
			&	H	& 1.6		& 1.171$^{+0.001}_{-0.001}$	& 1.									& 	-		& -			\\ \\
			&	K	& 2.2		& 0.875$^{+0.009}_{-0.005}$	& 0.747$^{+0.008}_{-0.004}$	& 1.			& -			\\ \\
			&	IRAC1	& 3.6	& 0.333$^{+0.032}_{-0.016}$	& 0.284$^{+0.027}_{-0.014}$	& 0.380$^{+0.041}_{-0.020}$	& 1.			\\ \\
			&	IRAC2	& 4.5	& 0.196$^{+0.020}_{-0.010}$	& 0.167$^{+0.017}_{-0.008}$	& 0.224$^{+0.025}_{-0.012}$	& 0.589$^{+0.002}_{-0.001}$	\\ \\
			&	IRAC3	& 5.8	& 0.132$^{+0.013}_{-0.007}$	& 0.113$^{+0.011}_{-0.006}$	& 0.151$^{+0.017}_{-0.008}$	& 0.396$^{+0.002}_{-0.001}$	\\ \\
			&	IRAC4	& 8		& 0.081$^{+0.008}_{-0.004}$	& 0.070$^{+0.007}_{-0.004}$	& 0.093$^{+0.011}_{-0.005}$	& 0.244$^{+0.002}_{-0.001}$	\\ \\
			&	WISE3	& 12	& 0.066$^{+0.007}_{-0.003}$	& 0.056$^{+0.006}_{-0.003}$	& 0.075$^{+0.009}_{-0.004}$	& 0.197$^{+0.002}_{-0.001}$	\\ \\
			&	WISE4	& 22	& 0.014$^{+0.002}_{-0.001}$	& 0.012$^{+0.001}_{-0.001}$	& 0.016$^{+0.002}_{-0.001}$	& 0.043$^{+0.001}_{-0.0003}$	\\ \\
			&	MIPS1	& 24	& 0.015$^{+0.002}_{-0.001}$	& 0.013$^{+0.001}_{-0.001}$	& 0.017$^{+0.002}_{-0.001}$	& 0.044$^{+0.0005}_{-0.0003}$	\\ \\
			&	IRAS	& 60	& 0.003$^{+0.0003}_{-0.0002}$ & 0.002$^{+0.0003}_{-0.0001}$	& 0.003$^{+0.0004}_{-0.0002}$	& 0.009$^{+0.0001}_{-0.00006}$	\\ \\					
	\hline			
	\end{tabular}
	\end{table*}
	
\section{\label{defparam} \cite{DraineLi07} parameters of the HRS galaxies.}

Even if this work is focused on the study of the dust properties of a subsample of gas rich galaxies from the HRS, we apply the SED fitting procedure presented in Section~\ref{fitresults} to all of the HRS galaxies.
First we removed the MIR stellar emission using the coefficients presented in Appendix~\ref{stell}.
In order to have fiducial corrected MIR flux densities, we impose the following criterion. 
If a galaxy has a corrected flux density lower than 2$\sigma$ at 8 or 12\microns, then this galaxy is removed from the sample.
At longer wavelengths, the contribution from the stellar populations to the IR emission is considered as negligible. 
		
270 out of the 322 HRS galaxies fulfill this criterion.
However, from these 270 galaxies, we remove 5 that were not detected in PACS and SPIRE bands, thus having no constrains on FIR and submm part of the SED: HRS\,90, HRS\,155, HRS\,240, HRS\,291 and HRS\,316.
We run our fitting procedure on 265 galaxies: 20 early-type galaxies and 245 late-type galaxies.

We present the results from the fits of the galaxies not analyzed in this work (20 early-type galaxies and 99 H{\sc i}-deficient galaxies) in Table~\ref{result_param_ETG_DEF}.
\tiny
\onecolumn{
\begin{longtable}{c c c c c c c c c c c c }
\caption{\label{result_param_ETG_DEF} Estimated parameters.} \\
\hline
\hline
HRS  & CGCG   &  VCC   & UGC  &  NGC  &  IC   & Type  & $U_{min}$ &  $q_{PAH}$ (\%) & $\gamma$ (\%) & $\log (L_{IR}/\rm{L_{\odot}})$ &  $\log (M_{dust}/\rm{M_{\odot}})$  \\
\hline
 \endfirsthead
\caption{continued.}\\  
\hline
\hline
HRS  & CGCG   &  VCC   & UGC  &  NGC  &  IC   & Type  & $U_{min}$ &  $q_{PAH}$ (\%) & $\gamma$ (\%) & $\log (L_{IR}/\rm{L_{\odot}})$ &  $\log (M_{dust}/\rm{M_{\odot}})$  \\
\hline
\endhead
\hline
\endfoot
       1       &    123035       &         -       &         -       &         -       &         -       &       pec       &            2.23   $\pm$            0.83       &            4.14   $\pm$            0.61       &            0.66   $\pm$            0.67       &            8.48   $\pm$            0.04       &            6.01   $\pm$            0.12       \\
       3       &     94026       &         -       &      5617       &      3226       &         -       &         E       &            2.38   $\pm$            0.41       &            4.46   $\pm$            0.27       &            0.18   $\pm$            0.09       &            8.49   $\pm$            0.02       &            6.02   $\pm$            0.06       \\
       4       &     94028       &         -       &      5620       &      3227       &         -       &        Sa       &            2.18   $\pm$            0.40       &            4.58   $\pm$            0.02       &            5.60   $\pm$            1.06       &           10.03   $\pm$            0.01       &            7.37   $\pm$            0.05       \\
       5       &     94052       &         -       &         -       &         -       &       610       &        Sc       &            2.64   $\pm$            0.41       &            4.05   $\pm$            0.42       &            0.37   $\pm$            0.29       &            9.23   $\pm$            0.03       &            6.67   $\pm$            0.05       \\
       6       &    154016       &         -       &      5662       &      3245       &         -       &        Sb       &            0.27   $\pm$            0.24       &            3.02   $\pm$            1.27       &            1.10   $\pm$            0.80       &            8.24   $\pm$            0.07       &            6.80   $\pm$            0.22       \\
\hline
\end{longtable}
}
\twocolumn
\normalsize

\section{\label{apcompparam} Comparison between the properties of the gas-rich galaxies and the output of DL07 models.}

\begin{figure*}
		\includegraphics[width=24cm,angle=-90]{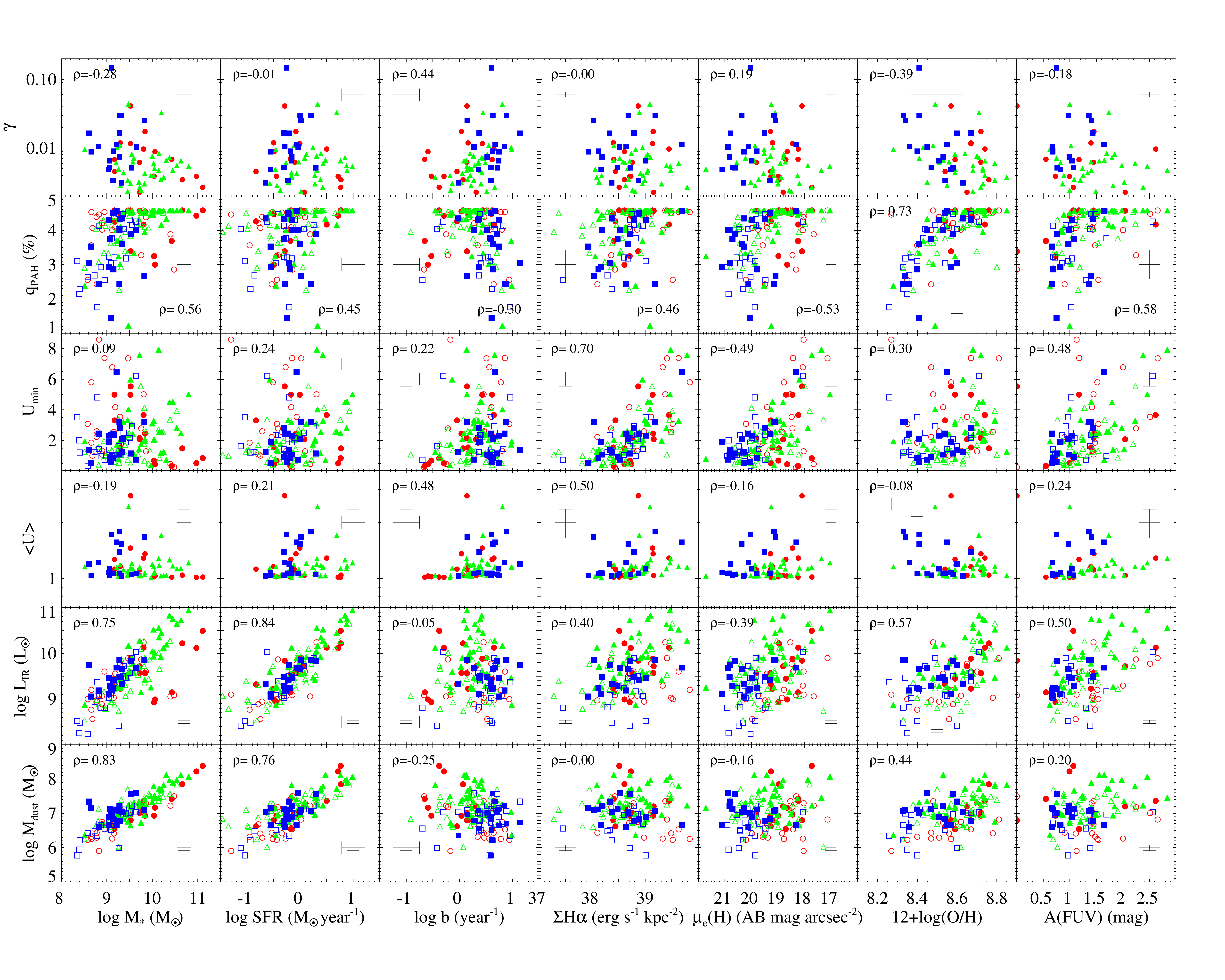}
  		\caption{ \label{totparam}Comparison between the output parameters of the \cite{DraineLi07} models and different physical variables. Galaxies are color-coded according their morphological type. In red: the Sa, Sab, Sb; in green: the Sbc, Sc, Scd, and in blue: the Sd, Im and BCD. The Spearman correlation coefficient of each relation is provided. Galaxies with a 24\microns\ measurement are represented by filled symbols, galaxies with 22\microns\ with empty symbols.}
	\end{figure*}

While we describe the main results of the comparison between the output parameters of DL07 and the properties of the gas-rich galaxy sample in Section~\ref{comp}, we describe here, for each output parameters, the relations with these properties.
These relations are presented in Figure~\ref{totparam}.
The Spearman coefficient is provided for every relation.
With a number of objects larger than 100, a correlation is expected to be real with a Spearman correlation coefficient larger than 0.40.

	\subsection{PAH emission}
	
	The fraction of PAH correlates with all of the parameters, except the birthrate parameter, with $\rho$ ranging from 0.45 to 0.73.
	The relation between the fraction of PAH, the metallicity and the stellar mass is described in Section~\ref{comp}.
	
	There is a weak relation between the fraction of PAH contributing to the total IR luminosity and the SFR ($\rho$=0.45).
	We would expect a tighter correlation as PAHs are often used as SFR indicators although many caveats on this assumption have been discussed in the literature \citep{Boselli04,Wu05,Calzetti07,Zhu08,Kennicutt09}.
	Indeed, the destruction of PAH in regions where the ISRF is too intense, such as PDRs, can affect the relation between the SFR and 8\microns\ luminosity.
	This leads to a weak anti-correlation with the birthrate parameter as well, and a moderate relation with the H$\alpha$ surface brightness that directly probes star formation.
	It seems that galaxies experiencing a star formation episode, i.e. with higher $b$, have lower $q_{PAH}$.
	Indeed, previous works found that the PAH emission is inhibited within star-forming regions relative to other star formation tracers \citep[e.g.][]{Helou04,Calzetti05,Bendo06,Perez-Gonzalez06,Bendo08,Gordon08}. 
	Another relation is found between $q_{PAH}$ and the H-band surface brightness, linked to what was previously noticed by \cite{Calzetti07} and \cite{Bendo08} who showed that 8\microns\ emission is also contributed by dust heated by the diffuse non-ionizing stellar component.
	A correlation is also found between the fraction of PAH and the FUV attenuation ($\rho$=0.58).
	However, the trend seen on the related panel is difficult to interpret, as the relation, if real, appears to be non-linear.

	\subsection{Relative contribution of PDR and diffuse regions}

	Here we only consider galaxies for which MIPS 24\microns\ measurements are available (filled symbols on Figure~\ref{totparam}) in order to have the strongest constraint on $\gamma$ (see Section~\ref{mocks}).
	There is a moderate relation between $\gamma$ and the birthrate parameter ($\rho=$0.44).
	When $\gamma$ increases, the contribution of the PDR to the emission of the IR SED increases.
	As $b$ is linked to the hardness of the UV radiation field, both quantities are correlated. 
	Although weak, these relations between $\gamma$ and $b$, and $\gamma$ and the metallicity are consistent as the most metal poor objects are also the most star forming.
	We also notice a trend between $\gamma$ and the metallicity, with a Spearman correlation coefficient of $\rho$=-0.39.
	The PDR contribution to the IR SED increases when the metallicity decreases.
	This confirms the tendency between $\gamma$ and the stellar mass already noticed from the shape of the IR SED in Figure~\ref{sedfit1}. 
	Furthermore, when $\gamma$ increases, the IR peaks widen (Figure~\ref{sedfit1}).
	These two points confirm the results of \cite{SmithDunne12} who performed a panchromatic analysis of the SED of a 250\microns\ selected sample of galaxies and found that low mass galaxies have broader IR peaks.
	This relation also implies that the IR SED of the most massive galaxies is dominated by the emission of the diffuse component.

	\subsection{Interstellar radiation field}

	The ISRF is characterized by the parameters $U_{min}$, which quantifies the ISRF of the diffuse stellar component, and $<U>$, the dust heating rate parameter, which is calculated from $U_{min}$, $\gamma$ and $U_{max}$ (fixed to 10$^6$).
	Relations between $<U>$ and the integrated properties of the galaxies are thus linked to the relations between $U_{min}$, $\gamma$ and these properties.
	
	The relation between the intensity of the diffuse ISRF and the H$\alpha$ and H-band surface brightnesses are discussed in Section~\ref{comp}.
	A weak correlation is observed between the birthrate parameter $b$ and $<U>$, with $\rho=$0.48.		
	The dust heating parameter $<U>$ provides a direct measurement of the mean interstellar radiation field of the galaxy.
	This confirms the results of \cite{Boselli10a} and \cite{Boselli12} who found a correlation between $b$ and the $S_{60}/S_{100}$ flux density ratio, sensitive to the dust temperature.
	
	\subsection{Infrared luminosity}
	
	An expected trend is seen between the $L_{IR}$ and $M_*$ as the most massive galaxies are also the most luminous as a scale effect \citep{Kennicutt90}.
	We recover as well the strong relation between the $L_{IR}$ and the SFR \citep{DevereuxYoung90,Devereux95,BuatXu96,Kennicutt09}.
	Indeed, the $L_{IR}$ is widely used as a proxy for the SFR \citep[e.g.,][]{ScovilleYoung83,BuatXu96,Kennicutt98,Kennicutt09,KennicuttEvans12}.
	
	We observe a moderate relation between the $L_{IR}$ and the metallicity ($\rho$=0.58).
	Finally, there is also a moderate trend between the $L_{IR}$ and the FUV attenuation ($\rho$=0.50).
	The energy absorbed in UV is re-emitted by the dust in IR, one would expect to find a tighter relation between these two properties.
	Indeed, \cite{SmithDunne12} found that the FIR/optical ratio increases with the $L_{IR}$ indicating that the galaxies with the higher IR luminosities are also the most obscured.
	However, by definition, $A(FUV)$ is the infrared to UV luminosity ratio.
	These two quantities are thus not independent. 
		
	\subsection{Dust mass}
	
	As expected, the dust mass and the stellar mass are tightly linked with a Spearman correlation coefficient of 0.83, as a scaling effect.
	There is also a strong correlation between the dust mass and the SFR.
	A weak global trend appears with the metallicity ($\rho$=0.44).
	Indeed the dust mass is larger in the most massive galaxies that are also the most metal-rich \citep{Tremonti04}.

	From Figure~\ref{totparam}, we conclude that all of the integrated properties, except the FUV attenuation, correlate moderately to strongly with at least one of the output of \cite{DraineLi07} models: the stellar mass to the fraction of PAH, the SFR to the fraction of PAH, the birthrate parameter moderately with $\gamma$ and $<U>$, $\Sigma H\alpha$ and $\mu_e(H)$ with $U_{min}$ and $q_{PAH}$, and the metallicity with $q_{PAH}$ and moderately with $\gamma$.
	All of these properties drive the shape of the IR SED.
	We confirm the results of \cite{Boselli12} who found that the metallicity, the intensity of the ionizing and non-ionizing radiation field, and the birthrate parameter are key parameters in the dust emission observed in the FIR.

\end{appendix}

\end{document}